\documentclass[a4paper,11pt]{article}

\pdfoutput=1 

\usepackage{jcappub} 

\usepackage[T1]{fontenc} 
\usepackage[section]{placeins}
\usepackage{graphicx}
\usepackage{booktabs}
\usepackage[normalem]{ulem}
\usepackage{aas_macros}
\useunder{\uline}{\ul}{}

\newcommand{\unit}[1]{\ensuremath{\, \mathrm{#1}}}

\newcommand{\be}{\begin{equation}}
\newcommand{\ee}{\end{equation}}
\newcommand{\bea}{\begin{eqnarray}}
\newcommand{\eea}{\end{eqnarray}}
\newcommand{\Msun}{M_\odot}
\newcommand{\Rsun}{R_{\odot}}
\newcommand{\Rstar}{R_{\star}}
\newcommand{\Mstar}{M_{\star}}
\newcommand{\Ncut}{N_{cutoff}}
\newcommand{\Mmax}{M_{max}}
\newcommand{\sigmav}{\langle\sigma v\rangle_{\mathrm{ann}}}

\title{Multiscatter capture of superheavy dark matter by Pop III stars}

\author[a,b,1]{Cosmin Ilie,\note{Corresponding author.}}
\author[a]{Saiyang Zhang}

\affiliation[a]{Department of Physics and Astronomy, Colgate University\\
13 Oak Dr., Hamilton, NY 13346, U.S.A.}
\affiliation[b]{Department of Theoretical Physics, National Institute for Physics and Nuclear Engineering\\
 Magurele, P.O.Box M.G. 6, Romania}
 
\emailAdd{cilie@colgate.edu}
\emailAdd{szhang@colgate.edu}

\abstract{If captured by the gravitational field of stars or other compact objects, dark matter can self-annihilate and produce a potentially  detectable particle flux. In the case of superheavy dark matter (\emph{$ m_{X} \gtrsim 10^{8} GeV $}), a large number of scattering events with nuclei inside stars are necessary to slow down the dark matter particles below the escape velocity of the stars, at which point the Dark Matter (DM) particle becomes trapped, or captured. Using the recently developed analytical formalism for multiscatter capture, combined with the latest results on the constraints of dark-matter-baryon scattering cross-section, we calculate upper bounds on the capture rates for superheavy dark matter particles by the first (Pop.~III) stars. Assuming that a non-zero fraction of the products of captured superheavy dark matter (SHDM) annihilations can be trapped and thermalized inside the star, we find that this additional heat source could influence the evolutionary phase of Pop.~III stars. Moreover, requiring that Pop.~III stars shine with sub-Eddington luminosity, we find upper bounds on the masses of the Pop.~III stars. This implies a DM dependent cutoff on the initial mass function (IMF) of Pop.~III stars, thus opening up the intriguing possibility of constraining DM properties using the IMF of extremely metal-poor stars. 
}

\begin{document}
\maketitle

\section{Introduction}
\label{sec:intro}

The first stars formed via the gravitational collapse of zero metallicity primordial baryonic gas clouds that contain H and He from big bang nucleosynthesis. The gas collapses on the gravity well provided by dark matter, at the center of mini-halos ($M_{halo}\sim 10^6 M_{\odot}$). Those DM halos form in the early universe via the gravitational growth of the initial density perturbations provided by cosmic inflation, and subsequent mergers, i.e. hierarchical structure formation. The emergence of the first stars marks the end of the ``dark ages'' of the universe. Using hydrodynamical numerical simulations of gas cloud collapse in the early universe, the following picture emerges: the first generation of stars (Pop.~III stars) have formed at redshifts $z=10 - 50$, when the universe was roughly 200 million years old. At that stage, the main cooling mechanism that allows the baryonic gas cloud to collapse towards the center of the halo and form a protostar is molecular hydrogen cooling. Collapse itself leads to heating of the protostellar gas cloud, and when a balance between heating and cooling is achieved, a protostar is formed (i.e. an object in hydrostatic equilibrium supported against further collapse by radiation and thermal gas pressure). This object continues to accrete material and become hotter and hotter, until eventually hydrogen burning has begun, and a zero age main sequence (ZAMS) zero metallicity star (a Pop.~III star) is formed. For reviews of the standard picture of the formation Pop.~III stars, see~\cite{Barkana:2000,Abel:2001,Bromm:2003,Yoshida:2006,Yoshida:2008,Loeb:2010,Bromm:2013,Klessen:2018}.

One important question is the following: can Dark Matter heating alter the formation or evolution of the first stars? This was first addressed by~\cite{Spolyar:2008dark}, for the case of self-annihilating weakly interacting particles (WIMPs) dark matter, the most promising -on theoretical grounds- DM scenario at the time. The authors found that given three generic assumptions -see details below- the collapse of the protostellar cloud can be halted by dark matter heating. This leads to a new phase of stellar evolution, a ``Dark Star (DS).'' The three conditions identified by~\cite{Spolyar:2008dark} that could lead to the formation of a DS are: (1) high DM density at the location of collapsing gas cloud, (2) some of the annihilation products can become trapped and thermalized inside the gas cloud/star, and (3) DM heating dominates over other heating or cooling mechanisms. Once the protostellar collapse has stopped, due to DM heating, a Dark Star is formed. This object will continue to accrete mass from the baryon cloud, and since it is typically cooler than a corresponding Pop.~III star, it can become much more massive, as feedback effects that stop the accretion process are directly proportional to the surface temperature. The dark star phase has been subsequently confirmed by~\cite{Iocco:2008}.

There are be two distinct sources for dark matter that can pile up and annihilate inside a star: (i) Adiabatic Contraction (AC) and (ii) capture. Adiabatic contraction~\cite{Blumenthal:1985} is a mechanism that is responsible for driving up the density of dark matter at the center of a collapsing gas cloud. Simply put, DM orbits shrink in response to the increase in the gravitational potential, and an increase in the baryon number density leads to an increase in the DM density. Capture is a mechanism via which dark matter particles can be trapped inside  compact baryonic objects. For the case of WIMP dark matter, this phenomenon was studied in the literature when the baryonic object is: the Sun~\cite{Press:1985}, the Earth~\cite{Gould:1987,Gould:1987resonant}, a dark star~\cite{Freese:2008cap,Iocco:2008cap,Sivertsson:2011}, a neutron star or a white dwarf~\cite{Bertone:2008}, etc. The case of strongly interacting superheavy DM capture by the sun was studied by~\cite{Albuquerque:2001}. For all of the above scenarios the physics is similar: if the orbit of a DM particle crosses the said object, there is a chance that collisions with the baryons will lead to the DM particle losing enough energy to become gravitationally trapped (i.e. whenever its speed becomes less than the escape velocity at the surface of the object).

Within the WIMP paradigm, the evolution and stellar structure of DSs has been extensively studied over the past decade. We summarize here some of the main results. Dark stars are born with masses $\sim 1 \Msun,$ for a large variety of WIMP masses, and grow to much larger masses via accretion~\cite{Freese:2008ds}. As alluded before, they are much puffier, and hence somewhat cooler than Pop.~III stars, with radii of about $10~\unit{AU}$ and surface temperatures of $\sim 10^4~\unit{K}$. A dark star is essentially an object in hydrostatic equilibrium that is supported against gravitational collapse by the pressure due mostly to the heat deposited inside the star via the thermalization of products of DM-DM annihilations happening inside the star. At a certain phase during the evolution of a DS, the adiabatically contracted DM reservoir will be depleted by annihilations. If one assumes spherical DM halos, this will happen when the DS has annihilated about $1~\Msun$ of DM.~\cite{Freese:2008ds,Spolyar:2009} found that this is sufficient to allow the DS to reach, via accretion, a mass of $\sim 10^3 \Msun,$ after about $300,000$ yr. At this stage a short contraction phase ensues, as now the DM annihilations are no longer efficient enough to prevent gravitational collapse. As the core temperature and density increase, nuclear fusion and/or DM capture can become efficient and prevent further collapse. Once a new equilibrium phase is reached, the DS enters the Zero Age Main Sequence (ZAMS).

N-body simulations indicate that DM halos are prolate-triaxial~\cite{Bardeen:1986,Barnes:1987,Frenk:1988,Dubinksi:1991}, which leads to a large part of the orbits passing arbitrarily close to the center, i.e. centrophilic orbits~\cite{Gerhard:1985,Meritt:1996,Meritt:1996b}. As such, the AC DM reservoir can be refilled efficiently, and hence the DS phase prolonged significantly~\cite{Freese:2010smds}. Within this ``extended AC'' phase, a DS could become supermassive, i.e. $M_{DS}\gtrsim 10^6\Msun$. A separate mechanism that could lead to the formation of a Supermassive Dark Star (SMDS) is capture of Dark Matter, if the ambient DM density is high enough. SMDSs are very bright and could be observed with the James Webb Space Telescope (JWST), as shown by~\cite{Freese:2010smds}. In view of their relatively cooler temperatures, compared to Pop.~III stars, one could also distinguish them from early galaxies formed by Pop.~III stars~\cite{Ilie:2012}. After the DM fuel runs out, a SMDS will quickly undergo a nuclear burning phase and then collapse to a Supermassive Black Hole (SMBH). We mention that in the standard picture, where the first stars are Pop.~III stars, the formation of the SMDS required to power the very bright observed high redshift quasars\footnote{For example J1342+0928~\cite{Banados:2018} is a quasar at $z=7.45$ that is powered by a behemoth BH with a mass of $8\times 10^8\Msun$} poses significant theoretical challenges. It requires either super-Eddington accretion or non-standard direct collapse to black holes (DCBH), two non-orthodox scenarios.  In contrast, SMDS, and the SMBH they will form, are a natural solution for this problem~\cite{Freese:2010smds,Banik:2019}.

Most of the work on Dark Stars has been done under the assumption of a thermal WIMP DM particles, i.e. thermal relics with masses in the $\unit{GeV}-\unit{TeV}$ range, generated in matter-antimatter collisions in the early universe. Considering a variety of thermal DM models~\cite{Gondolo:2010dmds} finds that in general this kind of models lead to the formation of DSs. However, at the LHC no particle has been identified as dark matter, while at the same time, direct detection experiments such as LUX and XENON have ruled out a large swath of the preferred parameter space for thermal WIMP DM. This places very tight constraints on the realization of the standard thermal relic WIMP paradigm. Therefore, non-thermal models started to become more and more relevant. Of those we mention only two: at the lower end of the mass spectrum one has the axions~\cite{Duffy:2009}, whereas WIMPZILLAs~\cite{Kolb:1999} are one example of superheavy dark matter (SHDM) particles. We note that there are many more DM particle models that are still within experimental bounds. For reviews on the state of particle dark matter, see~\cite{Bertone:2005,Abdallah:2015}. For a more general review on the state of dark matter in the universe, in view of all experimental constraints, see~\cite{Freese:2017dm}. 

It is therefore important to revisit the DS picture when considering non-thermal DM models. In this work we analyze if the evolution of Pop.~III stars could be altered by capture of SHDM particles with masses in the $10^8 - 10^{15}~\unit{GeV}$ range~\footnote{This range of masses for the Dark Matter candidate is representative for non-thermal relics produced during inflation, such as WIMPZILLAs~\cite{Kolb:1999}}. Additionally, we assume that DM annihilations is not preventing a Pop.~III star from forming and therefore we only consider its effects after the Pop.~III star has reached the zero metallicity zero age main sequence and DM capture can become important. In a future study we plan to investigate under what conditions, if any, heating due to annihilations of adiabatically contracted superheavy dark matter could halt the collapse of protostellar gas cloud prior to ignition of nuclear fusion, and therefore lead to the formation of a Dark Star instead of a Pop.~III star. Using the recently developed formalism for multiscatter capture of dark matter~\cite{Bramante:2017}, we calculate the capture rates and associated luminosity due to captured SHDM annihilations inside Pop.~III stars. Assuming a fiducial value of $\rho_X=10^{9}\unit{GeV/cm^3}$ for the ambient DM density, we find that the luminosity due to dark matter annihilations (DMAs) is typically orders of magnitude below the one due to nuclear fusion. For the most massive Pop.~III stars, that shine at or near the Eddington limit, the additional heat source would quickly destabilize them and lead to an unbound object. Therefore one possible effect of DMA is to impose an upper-bound on the masses of the Pop.~III stars. Since the DM heating is proportional to the ambient DM density, this upper bound is lower for higher values of $\rho_X$. As mentioned before the ambient dark matter density can be enhanced by many orders of magnitude near the center of a micro dark matter halo during the collapse of a protostellar gas cloud via a process called adiabatic contraction. For example, we find that, for an extreme case of $\rho_X=10^{18}\unit{GeV/cm^3}$, the Eddington limit places a bound on the masses of Pop.~III stars of a few $\Msun$. All of our results are upper bounds, obtained by using the assumption that the DM-nucleon scattering cross section ($\sigma_{n}$) has the highest possible value allowed by the exclusion limits from the latest direct detection experiments for strongly interacting massive particle (SIMP) dark matter obtained in~\cite{Kavanagh:2018}. 

This paper is organized as follows: in section~\ref{sec:multiscatter} we review the physics of dark matter capture, focusing on the main ingredients in the analytical formalism of multiscatter capture of dark matter introduced by~\cite{Bramante:2017}. In section~\ref{sec:rates} we calculate the capture rates and associated luminosities due to SHDM annihilations inside Pop.~III stars. Section~\ref{sec:masslimit} is dedicated to a discussion of one important effect of the additional heating from DM annihilations on Pop.~III stars: a dark matter dependent upper bound on Pop.~III stellar masses. We end with a discussion and conclusions in section~\ref{sec:conclusion}. 

\section{Capture of Dark Matter: brief review}
\label{sec:multiscatter}

As it transits a gravitationally bound object (e.g. star, earth, dark star, neutron star, etc) of radius $R_{\star}$, a dark matter particle would have collided with its constituents (e.g. nuclei) an average number of times given by:

\be\label{eq:navg}
N\approx n_T\sigma_n 2R_{\star}
\ee
Here $n_T$ is the average number density of collision targets (e.g. nuclei) inside the star, and $\sigma_n$ is the cross section of DM-nucleon scattering cross section. For the case of capture of weakly interacting dark matter particles with mass not much larger than $\sim 100~\unit{GeV}$ by the Sun~\cite{Press:1985} or the Earth~\cite{Gould:1987,Gould:1987resonant}, the average number of such collisions is always less than unity, in view of the small scattering cross sections. For instance the XENON1T experiment places the most recent upper bounds for a $100~\unit{GeV}$ WIMP on the spin independent nucleon DM interaction cross section at the order $\sigma_{SI}\lesssim 10^{-46}~\unit{cm}^2$~\cite{Aprilie:2018}, while for spin dependent cross section one has: $\sigma_{SD}\lesssim 10^{-41}~\unit{cm}^2$~\cite{Aprilie:2019}. Note that for a fixed capturing object mass ($M_{\star}$), the average number of collisions scales as $N\sim\sigma_n/R_{\star}^2 $. Even for the case of the most compact objects, neutron stars, the average number of collisions is much less than unity, given the above bounds on the scattering cross sections. Therefore for WIMPs, one needs not consider the multiscatter capture formalism. Instead, in order to calculate the capture rates, one can use the single scattering capture formalism of~\cite{Press:1985,Gould:1987,Gould:1987resonant}. 

Note that the experimental bounds via underground direct detection experiments are limited at large DM particle masses by the lower DM flux. Namely, the flux scales as: $\Phi\sim n_{X}\sim\rho_{X}/m_{X}\sim m_{X}^{-1}$. This leads to sensitivities that drop as $1/m_{X}$. In turn this amounts to constraints on the nucleon-DM scattering cross sections that scale as $\sigma_{bound}\sim m_{X}$, and therefore weaker bounds at high masses.~\cite{Albuquerque:2001} considers the regime of strongly interacting ($\sigma_n\sim 10^{-24}\unit{cm}^2$) superheavy ($m_{X}>10^{10}\unit{GeV}$) dark matter particles by the sun. In view of the large cross section, the mean free path of the DM particles inside the sun is very short ($\sim 10^{-2}\unit{cm}$), which amounts to a very large number of collisions each DM particle has, on average, as it traverses the Sun. After each collision, depending on the scattering angles, the factional energy loss by a DM particle is uniformly distributed in the following interval:
\be\label{eq:rangeloss}
0\leq\frac{\Delta E_i}{E_i}\leq \beta_+.
\ee
In the above equation, $E_i$ represents the energy the dark matter particle  of mass $m_{X}$ has before the $i$-th collision, $\Delta E_i$ is the energy it has lost due to the collision with a target nucleus of mass $m_{n}$, and $\beta_{\pm}\equiv 4m_{X}m_{n}/(m_{X}\pm m_{n})^2$. Therefore the average fractional energy loss per collision is, assuming equal probability for all values in the kinematically allowed range of equation~\ref{eq:rangeloss}, just half of the upperbound  (i.e. $\beta_+/2$). The most efficient energy transfer happens when $m_{X}=m_{n}$, as expected. In the case of SHDM ($m_{X}\gg m_{n}$) one can see that the average fractional energy loss becomes $\Delta E/E\approx (1/2)(m_{n}/m_{X})\ll 1$. Again, this is to be expected. By analogy, a bowling ball would lose a very small fraction of its energy after a collision with a ping pong ball. Therefore, only DM particles at the low end of the velocity distribution can be captured in one collision. For most SHDM particles, it would take much more than one collision to slow down below the escape velocity. This is the case considered by~\cite{Albuquerque:2001}, where analytic formulae for capture rates of a class of strongly interacting SHDM known as SIMPZILLAs can be found. However, experimental constraints on SHDM-nucleon scattering cross sections have closed the window on such DM models~\cite{Albuquerque:2010closing,Kavanagh:2018}, and as such the formalism developed in~\cite{Albuquerque:2001} has little practical use for DM capture physics.  Capture of DM in the intermediary regime, where the average number of collisions it takes for a DM particle to be trapped is of order unity, can be treated using the formalism recently developed by~\cite{Bramante:2017}. This regime will be the one relevant for this work. In the reminder of this section we give a brief review of the main ingredients from~\cite{Bramante:2017} that we will subsequently use in this work.       

There are two factors that control the rate of capture: the flux ($F$) of DM particles through the star and the probability ($\Omega$) that each DM particle crossing will lose enough energy via collisions such that it becomes bound by the gravitational field of the star. For all the regimes described above, the schematic differential capture rate is given as~\cite{Press:1985,Gould:1987,Gould:1987resonant,Bramante:2017}:
\begin{align}\label{eq:DiffCapture}
\dfrac{dC}{dVd^{3}u}=dF(n_{X},u,v_{star},v_{esc}^{halo}) \Omega (n_{T}(r), \omega (r), \sigma_{n} , m_{n} , m_{X}), 
\end{align}       
where $u$ represents the DM velocity far from the gravitational field of the star, $v_{star}$ is the relative velocity of the star with respect to the halo, $v_{esc}^{halo}$ denotes the escape speed of the DM halo, $n_{T}(r)$ is the number density of scattering targets inside the star at a radius $r$ from the center, $w(r)^{2} =u^{2}+v_{esc}(r)^{2}$ is the velocity of the dark matter particles inside the star, with $v_{esc}(r)^{2}\equiv 2GM_{\star}(r)/r$. Since Pop III stars form usually at the center of DM halos, we can assume no relative motion of the star in the dark matter halo($ v_{star}\rightarrow 0 $). Additionally, the following assumptions are made: a very large escape speed for dark matter halo($ v_{esc}^{halo} \rightarrow \infty $), a uniform density  for the star, and a fixed escape velocity ($ v_{esc}(r) = v_{esc}(R_{\star})$) where $R_{\star}$ is the radius of the star. 
The total capture rate corresponding to each dark matter particle mass($ m_{X} $) is
\begin{align}\label{eq:Ctot}
C_{tot}(m_{X}) =  \sum_{N=1}^{\infty} C_{N},
\end{align}  
where $C_{N}$ is the capture rate after exactly N scattering events. The optical depth is conveniently defined as: $\tau\equiv n_T\sigma_{n} (2R_{\star})$, with $n_T$ representing the average number density of targets the DM particle can collide with as it traverses the star. In view of equation~\ref{eq:navg}, this also represents the average number of collisions the DM particle will experience as it traverses the star. In addition to the DM flux, there are two separate factors at play when one calculates $C_N$: the probability that a DM particle with optical depth $\tau$ will actually participate in $N$ scatterings ($p_N(\tau)$), and the probability that the velocity of the DM particle drops below the escape speed at the surface of the star after exactly N collisions ($g_N(w)$). Taking into account the geometry of the different possible incidence angles,~\cite{Bramante:2017} defines $p_{N}(\tau) = (2/N!)\int_{0}^{1} \,dy\, ye^{-y\tau}(y\tau)^{N}$. We note that the integral can be performed analytically, with the following result which we will use throughout our calculations:

\begin{align}\label{eq:pN}
p_{N}(\tau) = \dfrac{2}{\tau^{2}} \left(N+1-\dfrac{\Gamma(N+1,\tau)}{N!} \right).
\end{align}
We next consider $g_{N}(w)$. The initial kinetic energy of dark matter particle as it enters the stars is: $ E_{0}=m_{X}w^{2}/2 $. The energy loss after each elastic collision follows simple kinematics rules as $ \bigtriangleup E = z\beta_{+}E_{0}$, where $ z \in [0,1] $ is a kinematic variable related to the scattering angle, and  $\beta_{\pm}\equiv 4m_{X}m_{n}/(m_{X}\pm m_{n})^{2}$. Therefore, the kinetic energy after one collision reduces to $E_{i}=(1-z_{i}\beta_{+})E_{i-1}$ and the velocity of the DM particle becomes $v_{i}=(1-z_{i}\beta_{+})^{1/2}v_{i-1}$. In our case, since $ m_{X}\gg m_{n} $ we can approximate $ \beta_{+}\approx 4m_{n}/m_{X}$. After $ N $ scatters, the velocity and energy for a dark matter particle becomes\cite{Albuquerque:2001,Bramante:2017}:
\begin{align}
E_{N}= \prod_{i=1}^{N}(1-z_{i}\beta_{+})E_{0},\hspace{1cm}v_{N}= \prod_{i=1}^{N}(1-z_{i}\beta_{+})^{1/2}w.
\end{align}
Considering the capture condition ($v_N<v_{esc}$) and taking into account the different possible paths a DM particle can trace inside the star while undergoing $N$ scatterings (i.e. different scattering angles for each collision), one can express the probability $g_N(w)$ as~\cite{Bramante:2017}:
\begin{align}\label{eq:gw}
g_{N}(w)=\intop\nolimits_{0}^{1}dz_{1}\intop\nolimits_{0}^{1}dz_{2}\cdot\cdot\cdot\intop\nolimits_{0}^{1}dz_{N}\Theta \left(v_{esc}\prod_{i=1}^{N}(1-z_{i}\beta_{+})^{-1/2}-w \right).
\end{align}
The $\Theta \left(v_{esc}\prod_{i=1}^{N}(1-z_{i}\beta_{+})^{-1/2}-w \right)$ factor amounts to the probability of capture after a path through the star described by $N$ collisions, each with scattering angle determined by $z_i$. For the case of Spin Independent cross sections, considered in this work, one can further assume no preferred scattering direction (i.e. cross section for nucleon-DM scattering is independent of scattering angle). Therefore the average value of $z_i$ can be assumed to be equal to 1/2, leading to a simplified form for $g_N(w)$~\cite{Bramante:2017}: 
\begin{align}\label{eq:gwsimp}
g_{N}(w)=\Theta \left(v_{esc}\prod_{i=1}^{N}(1-z_{i}\beta_{+})^{-1/2}-w \right).
\end{align} 
In order to calculate the capture rate after exactly $N$, scattering events one needs to multiply the particle rate through the surface of the star with the probability that a DM particle will be captured after $N$ collisions (i.e. the product $p_N(\tau)*g_N(w)$). The capture rate after exactly $N$ collisions ($C_N$) can be expressed as the following phase space integral~\cite{Bramante:2017}:
\begin{align}\label{eq:CN}
C_{N} = \pi R^{2} p_{N}(\tau) \int_{v_{esc}}^{\infty}f(u)\dfrac{dw}{u^{2}} w^{3}g_{N}(w),
\end{align}
where $f(u)$ is the DM velocity distribution. The assumption that the star is not moving with respect the dark matter halo is implicit. The integral can be analytically evaluated assuming a Maxwellian distribution with an average speed $\overline{v}$ and using the probability $g_N(w)$ from equation~\ref{eq:gwsimp}. For the full result see~\cite{Bramante:2017}. Under the following two assumptions: (i) $v_{esc}\gg \bar{v}$ and (ii) $m_{X}\gg m_{n}$ (which are always valid for the cases we will consider in this work) the following simplified expression for $C_N$ can be obtained~\cite{Bramante:2017}:
\bea\label{eq:CNsimpl}
C_{N} &=&\sqrt{24 \pi} p_{N}(\tau)G\dfrac{\rho_{X}}{m_{X}}M_{\star}R_{\star}\dfrac{1}{\overline{v}}\left[ 1-\left(1-\dfrac{2A_{N}^{2}\overline{v}^{2}}{3v_{esc}^{2}} \right)e^{-A_{N}^{2}} \right];\\
A_{N}^{2}&\equiv&\dfrac{3 Nv_{esc}^{2}m_{n}}{\overline{v}^{2}m_{X}},
\eea
where $\rho_{X}$ is the DM density, and $M_{\star}$ and $R_{\star}$ are the mass and radius of the star, $m_{n}$ mass of nucleons in the star, $\overline{v}$ the dispersion velocity of DM, $v_{esc}$ the escape velocity at the surface of the star.

In the next section we will implement a numerical procedure to calculate the infinite series in equation~\ref{eq:Ctot} and use it for the case of SHDM being captured by Pop.~III stars. It is worth emphasizing that a constant density star is also assumed. Relaxing this assumption, and considering a variable density star, leads to an increase in the capture rates, as pointed out in~\cite{Bramante:2017}. With this in mind, all the bounds we compute in this paper should be viewed as conservative. In particular, this applies to the upper bounds on the total capture rates from section~\ref{sec:rates} and the associated Pop.~III stellar mass limits obtained in section~\ref{sec:masslimit}.

\section{Superheavy dark matter capture by Pop.~III stars}
\label{sec:rates}
In this section, we obtain upper bounds on capture rates and associated DM-DM annihilation luminosity in the case of superheavy dark matter capture by Pop.~III stars with mass ranging from a few to a thousand solar masses. As mentioned in section~\ref{sec:intro}, the first stars formed at the center of  dark matter mini-halos ($M_{halo}= 10^{5} - 10^{6} M_{\odot}$) at redshifts of  $z=10 - 50$. We assume that the  density profile of the dark matter halo follows the Navarro-Frenk-White\cite{Navarro:1997} profile:
\begin{align}\label{eq:NFW}
    \rho_{halo}=\dfrac{\rho_{0}}{\dfrac{r}{r_{s}}\left( 1+\dfrac{r}{r_{s}}\right)^{2}},
\end{align}
where $\rho_{halo}$ is the density of dark matter halo, $\rho_{0}$ the density at the center of the DM halo,$r$ the distance away from the center of the DM halo, and $r_{s}$ is the scale radius. And the central density $\rho_{0}$ is related to the critical density $\rho_{crit}$ at given redshift $z$, via
\begin{align}\label{eq:rho0}
\rho_{0}=\rho_{crit}(Z)\dfrac{200}{3}\dfrac{c^{3}}{\ln(1+c)-c/(c+1)},
\end{align}
where $c\equiv r_{vir}/r_{s}$ is the concentration parameter and $r_{vir}$ is the scale radius. From the virial theorem, the dispersion velocity $\overline{v}$ of DM inside the halo is
\begin{align}\label{eq:dispv}
\langle \overline{v}^2 \rangle = \dfrac{\overline{W}}{M_{halo}},
\end{align}
where
\begin{align}\label{eq:W}
    W=-4\pi G \int \rho_{halo}M_{halo}(r)rdr
\end{align}
is the gravitational potential of the dark matter halo.
\par
\begin{figure}[!htb]
\centering
\includegraphics[width=0.7\textwidth]{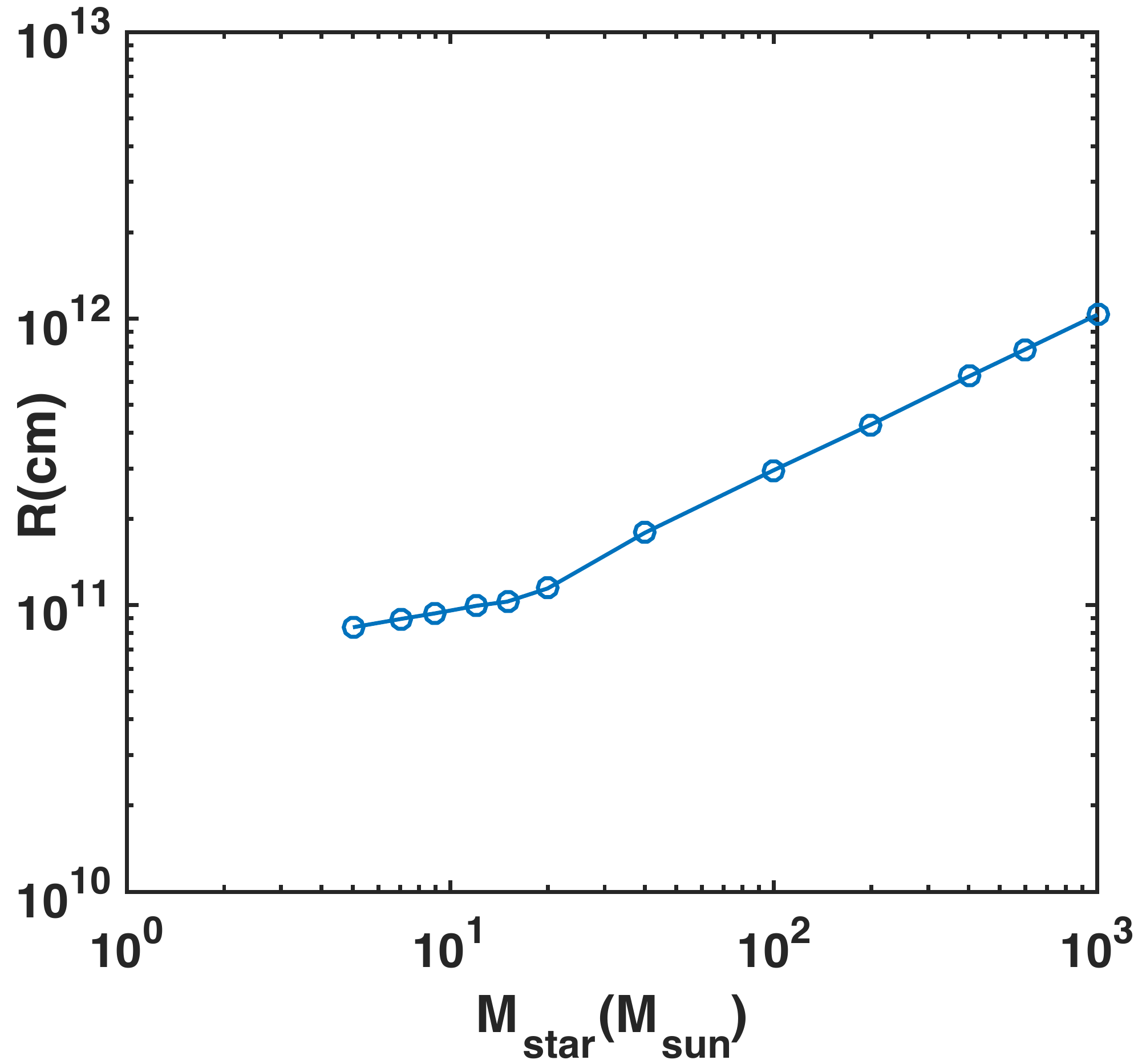}
\caption{\label{fig:RvsM} Stellar radius versus stellar mass for Pop.~III stars. We find that there are two distinct regimes, and therefore two different ``homology'' scaling relations can be applied. For stars with mass $M_{\star}\lesssim 20\Msun$ one gets $R_{\star}\propto M_{\star}^{0.21}$, whereas for more massive stars $R_{\star}\propto M_{\star}^{0.56}$}.
\end{figure}

 
We adopt here the same parameters for the NFW profile describing the DM halo where the first stars form as those used in~\cite{Freese:2008cap}, who studied the capture of WIMPs by Pop.~III stars. Namely, the following ranges: concentration parameter $c= 1 - 10$, redshift $z = 10 - 50$, and $r_s = 15 - 100~\unit{pc}$. For the DM dispersion velocity one gets values ranging from $\overline{v}= 1 - 15~\unit{km/s}$. Therefore we adopt the following fiducial value: 
\be\label{eq:vbarfidu}
\overline{v}=10~\unit{km/s}.
\ee
For the DM density at the location of the star (i.e. the  center of the DM halo), we assume $\rho_{X}=\rho_0=10^9~\unit{GeV/cm^3}$. From equation~\ref{eq:CNsimpl} using numerical values and substituting $p_N(\tau)$ from equation~\ref{eq:pN} we obtain the following scaling relation: 
\begin{multline}\label{eq:CNscaling}
C_{N}\simeq1.35\times10^{43}s^{-1}\left( N+1-\dfrac{\Gamma(N+2,\tau)}{N!}\right)  \left[ 1-\left(1-2\times 10^{-8}N \left( \dfrac{10^{8}GeV}{m_{X}}\right) \right)e^{-A_{N}^{2}} \right]\\\left( \dfrac{10^{8}GeV}{m_{X}}\right)\left( \dfrac{1.26\times10^{-40}cm^{2}}{\sigma_{n}}\right)^{2}  \left( \dfrac{10km/s}{\overline{v}}\right)\left( \dfrac{\rho_{X}}{10^9 GeV /cm^{3}}\right) \left( \dfrac{M_{\odot}}{\Mstar}\right)\left( \dfrac{\Rstar}{R_{\odot}}\right)^{5}, 
\end{multline} 
with the following approximate numerical scaling relationships for the average collision number ($\tau$) and the exponential factor ($A_N^2$):
\bea
\label{eq:AN}
A_{N}^{2} &\simeq& 1.10\times 10^{-4} N \left( \dfrac{v_{esc}}{618km/s}\right)^{2} \left( \dfrac{10km/s}{\overline{v}}\right)^{2} \left( \dfrac{10^{8}GeV}{m_{X}}\right),\\ 
\tau &\simeq& 1.10\times 10^{-5}\left( \dfrac{\sigma_{n}}{1.26\times10^{-40}cm^{2}}\right)\left( \dfrac{R_{\odot}}{\Rstar}\right)^{2}\left( \dfrac{\Mstar}{M_{\odot}}\right).\label{eq:tauscaling}
\eea 
Note that one could use mass-radius ``homology'' relationships to eliminate the $R_*$ dependence in equations~\ref{eq:CNscaling} and~\ref{eq:tauscaling}. For example, for massive ($M*\gtrsim 100\Msun$) Zero Age Main Sequence (ZAMS) stars of very low metallicity, $\Rstar\propto \Mstar^{5/11}$~\cite{Bromm:2001}. Also, see figure~\ref{fig:RvsM} and equations~\ref{eq:homologylow} and~\ref{eq:homologyhigh}, where we obtain the homology mass-radius relations for our set of Pop.~III stars.

At first glance it seems that the capture rate in equation~\ref{eq:CNscaling} is proportional to $\sigma_{n}^{-2}$, which is contrary to expectation. A larger scattering cross section should, in principle, lead to a larger capture probability, and hence a larger rate of particles captured. This is certainly the case for the single scatter capture, where $C\propto\sigma_n$. The  $\sigma_{n}^{-2}$ factor comes from $p_N(\tau)\propto 1/\tau^2\propto1/\sigma_{n}^2$. However, $p_N(\tau)$ contains the additional factor: $\left( N+1-\dfrac{\Gamma(N+2,\tau)}{N!}\right)$, which also has a $\tau$ (and hence $\sigma_n$) dependence. It is straightforward to show that in the case of $\tau \ll 1$:
\be\label{eq:GammaSeries}
N+1-\dfrac{\Gamma(N+2,\tau)}{N!}\approx\frac{\tau^{N+2}}{N!(N+2)}+\mathcal{O}(\tau^{N+3}).
\ee    
As expected, for $\tau \ll 1$, which is to say when the dark matter particle has a very large (compared to $R_{\star}$) mean free path through the star, $C_N$ will quickly vanish for  $N>1$. So, only $C_1$ is relevant, or, in other words the multiscatter formalism, in the appropriate limit, recovers naturally the single scatter formalism of~\cite{Press:1985,Gould:1987,Gould:1987resonant}. Note that for low mass (WIMP-like) dark matter capture, the exponential factor $A_N^2\gg 1$ and the entire term in the square brackets of equation~\ref{eq:CNscaling} reduces to one~\cite{Freese:2008cap}. Therefore, isolating the scaling of $C_1$ with DM parameters we get: $C_1\propto \sigma_n \rho_X/m_X,$ as expected and predicted by the single scatter capture analytic formulae of~\cite{Gould:1987resonant}.  

For the remainder of this work, we will adopt the XENON1T 2018 bounds~\cite{Aprilie:2018} on spin independent dark mater nucleon scattering cross section. Since the first stars are mostly ionized hydrogen  (nucleons), this bound applies to the cross section of the scattering of DM with the targets inside the first stars.  Therefore, all of our results on the capture rates and associate quantities such as DM-DM luminosity are upper bounds. Specifically, by fitting the XENON1T one year exposure upper limits in the $10^{2}\unit{GeV}\leq m_{X}\leq 10^{15}\unit{GeV}$, we get the following linear dependence:
\be\label{eq:signbound} 
\sigma_n(m_X)\lesssim 1.26\times10^{-40}\left(\frac{m_{X}}{10^8\unit{GeV}}\right).
\ee
We emphasize that this dependence is not due to any particle physics interaction; it is just a reflection of the fact that at large mass the sensitivity of direct detection experiments are limited by the DM flux. For theoretical expectations of the $\sigma_n$ dependence with $m_{X}$ for a large variety of DM models see~\cite{Jungman:1996,Ellis:2000,Cheng:2002}.

In the calculation of the capture rates we will need parameters -radii and masses- for the stars of interest (i.e. Pop.~III stars). As those are not yet observed, and it is unlikely even JWST will be able to observe them in isolation, so one has to resort to numerical simulations. We adopt models from~\cite{Iocco:2008}, for stars with mass in the $5~\Msun - 600~\Msun$ range. In addition we include a $1000\Msun$ Pop.~III star~\cite{Schaerer:2002,Ohkubo:2009}. We have checked that all models we considered are consistent with more recent simulations, such as those given in~\cite{Windhorst:2019}, for example. See table~\ref{table:PopIII} for a summary of the parameters of interest for the Pop.~III models we use.
\begin{table}[h!]
\centering
\begin{tabular}{@{}llll@{}}
\toprule
 $M_{\star}(M_{\odot})$  &$R_{\star}(R_{\odot})$  & $V_{esc}(V_{\odot})$  &$L_{\star}(L_{\odot})$\\ \midrule
      
       5    & 1.203 & 2.06 & $8.41\times 10^{2}$ \\
       7    & 1.286 & 2.36 & $2.39\times 10^{3}$  \\ 
       9    & 1.347 & 2.61 & $5.07\times 10^{3}$\\
       12   & 1.43 & 2.93  & $1.18\times 10^{4}$ \\
       15   & 1.48 & 3.22  & $2.19\times 10^{4}$ \\
       20   & 1.65 & 3.52  & $5.00\times 10^{4}$\\
       40   & 2.57 & 3.98  & $2.86\times 10^{5}$\\
       100  & 4.25 & 4.90  & $1.45\times 10^{6}$\\
       200  & 6.14 & 5.76  & $3.97\times 10^{6}$ \\
       400  & 9.03 & 6.72  & $9.89\times 10^{6}$\\
       600  & 11.24 & 7.38 & $1.61\times 10^{7}$ \\
       1000 & 14.83 & 8.29 & $3.50\times 10^{7}$\\\midrule 
\end{tabular}
\caption{Stellar mass ($M_{\star}$), radius ($ R_{\star} $), surface escape velocity ($V_{esc}$), and luminosity ($L_{\star}$) in solar units, for the Pop.~III models of~\cite{Iocco:2008} and~\cite{Ohkubo:2009} we consider in this paper.}
\label{table:PopIII}
\end{table}

Our first aim is to calculate $C_{tot}(m_{X})$, as given by equations~\ref{eq:Ctot} and~\ref{eq:CNsimpl}. In principle, the series defining the total capture rate is infinite, but in view of $p_{N}$ decreasing for large $N$, the series converges after a number of steps $N_{cut}$ that is  $\tau$ dependent. Defining the partial sum $C_{tot,N}\equiv\sum_{i=1}^{N}C_i$, our convergence criterion is the following: 
$$\vert (C_{tot,N_{cutoff}+1}/C_{tot,N_{cutoff}})-1\vert\leq 0.001.$$
We code this criterion  when estimating numerically the total capture rate $C_{tot}$ for each case we consider. It is important to note that for all the Pop.~III models we considered $A_N^2\ll 1$, even for the largest values of $N$. In view of equation~\ref{eq:AN}, this is due to the large values of $m_X$, and the fact that $p_N(\tau)$ drops to zero quickly after $N>\tau$, so values of $N\gtrsim1000$ are never needed. In fact, the largest value for $\Ncut$ is $818$, corresponding to the capture of $10^{15}\unit{GeV}$ dark matter particles by a $20\Msun$ Pop.~III star. Note that in the $A_N^2\ll 1$ and $\overline{v}\ll v_{esc}$ limits the term in the square brackets of the equations defining $C_N$ can be Taylor expanded. Keeping only the leading order terms we find:
\be\label{eq:Sqbrapprox}\nonumber
\left[ 1-\left(1-\dfrac{2A_{N}^{2}\overline{v}^{2}}{3v_{esc}^{2}} \right)e^{-A_{N}^{2}} \right]\approx A_N^2\approx1.10\times 10^{-4} N\left(\frac{10^8\unit{GeV}}{m_X}\right)\left(\dfrac{10km/s}{\overline{v}}\right)^2\left(\frac{M_{\star}}{\Msun}\right)\left(\frac{\Rsun}{R_{\star}}\right).
\ee
When taking this approximation into account, $C_N$ becomes: 
\bea \label{eq:CNscalingsimpl}
C_{N}&\simeq& 1.48\times10^{39}s^{-1}\left( N+1-\dfrac{\Gamma(N+2,\tau)}{N!}\right) N\left(\dfrac{10^{8}GeV}{m_{X}}\right)^2\nonumber\\
&&\left( \dfrac{1.26\times10^{-40}cm^{2}}{\sigma_{n}}\right)^{2}  \left( \dfrac{10km/s}{\overline{v}}\right)^3\left( \dfrac{\rho_{X}}{10^9 GeV /cm^{3}}\right) \left( \dfrac{R_{\star}}{R_{\odot}}\right)^{4}
\eea
Note here one important distinction, when compared to the low mass case (WIMPs), where $C_1\propto\sigma_n\rho_X/m_X$. For SHDM, in the case of $\tau\ll 1$, using the expansion from equation~\ref{eq:GammaSeries} we get $C_1(SHDM)\approx \sigma_n\rho_X/m_X^2$. The extra factor of $1/m_X$ suppression comes from the different way the term in the square brackets of equation~\ref{eq:CN} scales with mass: $\propto m_X^0$ for WIMPs vs. $\propto 1/m_X$ for SHDM, as previously explained.     

\begin{figure*}[!htb]
\begin{center}$
\begin{array}{c} 
\includegraphics[width=0.7\textwidth]{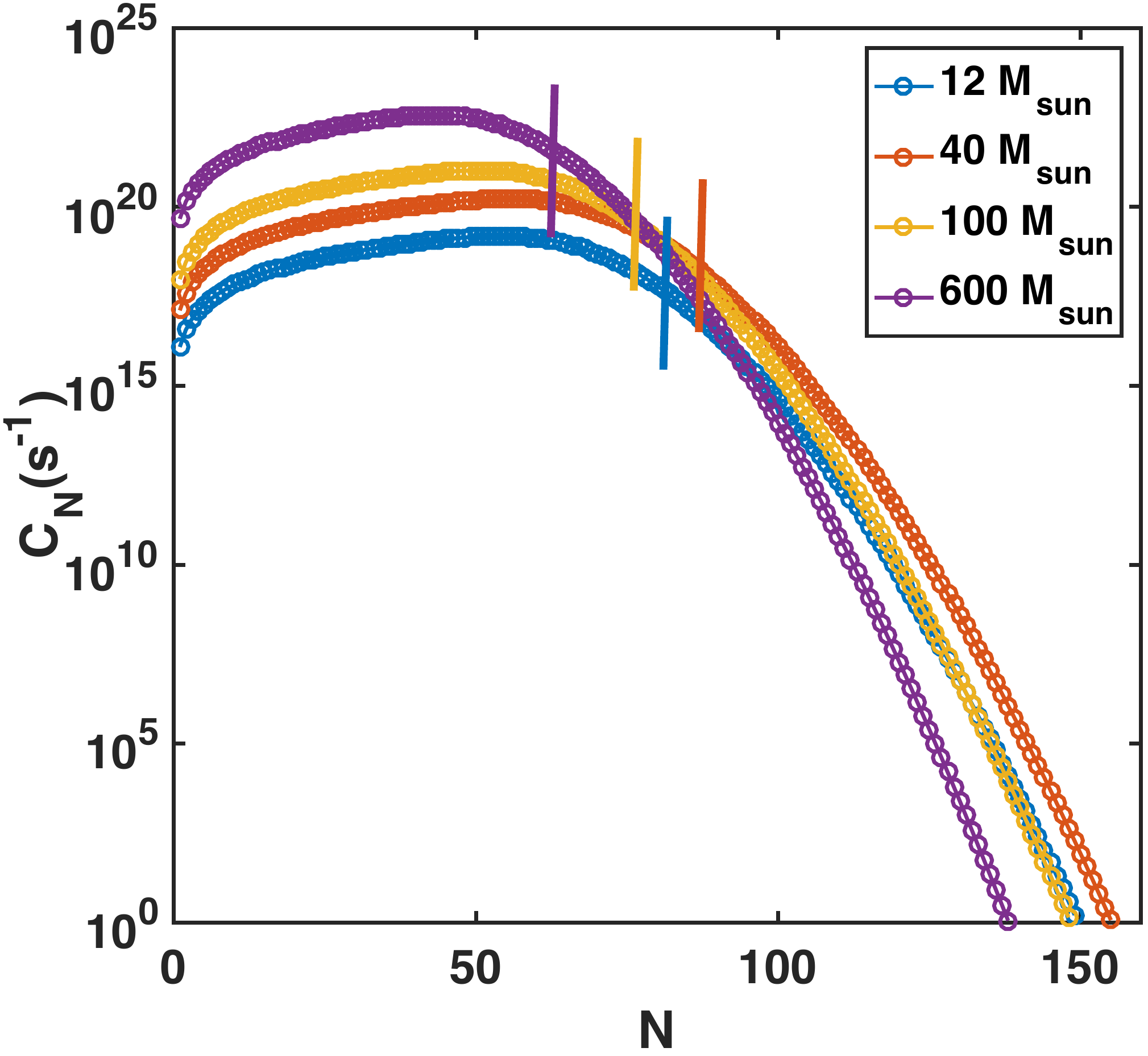}
\end{array}$
\end{center}
\caption{Capture rates for SHDM ($m_{X}=10^{14}\unit{GeV}$) after exactly $N$ collisions ($C_N$) for Pop.~III stars of  various masses. Note that the maximum value of $C_N$ increases with stellar mass. The bold vertical lines correspond to the points where the cutoff criterion is satisfied, i.e. $N=N_{cutoff}$. Unless otherwise specified, we always assume $\rho_X=10^9\unit{GeV/cm^3}$.}
\label{fig:CNvsN}
\end{figure*} 

In figure~\ref{fig:CNvsN}  we plot the coefficients $C_N$ (capture rate after exactly $N$ scatterings) calculated for the case of $m_{X}=10^{14}\unit{GeV}$ as a function of $N$ for several Pop.~III models. As expected, as large $N$ the rate $C_N$ drops significantly, approaching zero very fast after a certain threshold ($N_{cutoff}$). This is due to the decrease in the probability of capture if the number of scatterings ($N$) exceeds by a significant margin the average number of scatterings ($\tau$). In addition we note that the maximum value for $C_N$ increases with an increase in the mass of the star. This can be explained in the following way: $C_N\propto R_{\star}^4$ (see equation~\ref{eq:CNscalingsimpl}). Assuming a homology scaling relation $R_{\star}\propto M_{\star}^{p}$ we get $C_N\propto M_{\star}^{4p}$. The exponent is positive for any values of $p>0$. Numerically we find that the mass radius dependence can be well fit in the following way (see figure~\ref{fig:RvsM}): 
\bea
\frac{\Rstar}{\Rsun}&\simeq& 0.84\left(\frac{\Mstar}{\Msun}\right)^{0.21}, {\mathrm{when }}\, \Mstar\lesssim 20\Msun \label{eq:homologylow}\\
\frac{\Rstar}{\Rsun}&\simeq& 0.32\left(\frac{\Mstar}{\Msun}\right)^{0.56}, {\mathrm{when }}\, \Mstar\gtrsim 20\Msun\label{eq:homologyhigh}
\eea
Going back to figure~\ref{fig:CNvsN}, we point out that for the higher mass stars ($M_{\star}\gtrsim 20\Msun$) the value of $N_{cutoff}$ has a mild inverse dependence with the mass of the star, i.e. higher mass stars have lower $N_{cutoff}$. This is to be expected, as
\be\label{eq:NcutMstar}
N_{cutoff}\propto N_{average}\propto n_TR_{\star}\propto\frac{M_{\star}}{R_{\star}^2}\propto M_{\star}^{1-2p}.
\ee
For the higher mass stars we expect $N_{cutoff}\propto M_{\star}^{-0.12}$ in view of the numerically obtained value of $p=0.56$. We also confirmed this result using numerical fitting. For the lower mass stars, based on the same type of analysis, we have $N_{cutoff}\propto M_{\star}^{0.6}$.

\begin{figure}[!htb]
\centering 
\includegraphics[width=0.7\textwidth]{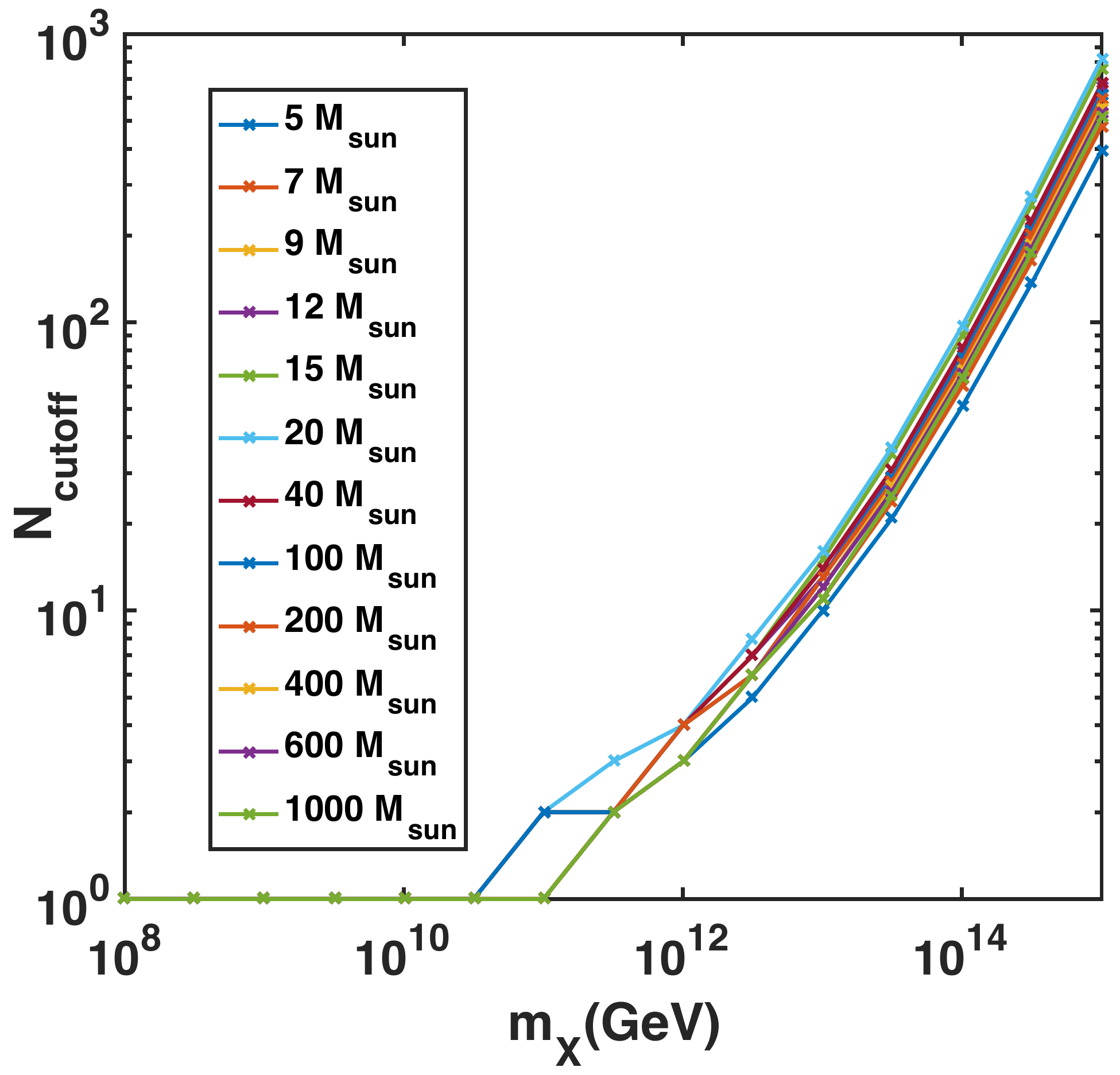}
\caption{\label{fig:Ncut} DM mass dependence of $N_{cutoff}$ for all of the Pop.~III models we consider. Note that for masses below $m_{X}\lesssim 10^{11}\unit{GeV}$ the cutoff is $N_{cutoff}=1$ and that at high masses $N_{cutoff}\propto m_{X}$.}
\end{figure}

Next we investigate the dependence of $N_{cutoff}$ with the mass of the dark matter particle. We expect a transition to happen whenever $\tau(m_X)\sim 1$ is reached. The exact value of $m_X$ where this transition takes place depends on the mass and radius  of the star (see equation~\ref{eq:tauscaling}), via the following combination $M_{\star}/R_{\star}^2$. As pointed out before, for Pop.~III stars with $M_{\star}\gtrsim 20\Msun$, this converts into a very mild dependence with mass of the star: $M_{\star}^{-0.12}$. For all cases considered, this transition happens when $m_X\sim 10^{12}\unit{GeV}$. For lower $m_X$, when $\tau\leq 1$, we recover the single scattering formalism, and therefore $\Ncut\approx 1$. As $\tau\propto\sigma_n$, and the upper bound on the scattering cross section scales linearly with $m_X$, we expect a linear dependence of $\Ncut$ with $m_X$. All those trends can be confirmed in figure~\ref{fig:Ncut}. 

\begin{figure}[!htb]
\centering 
\includegraphics[width=0.7\textwidth]{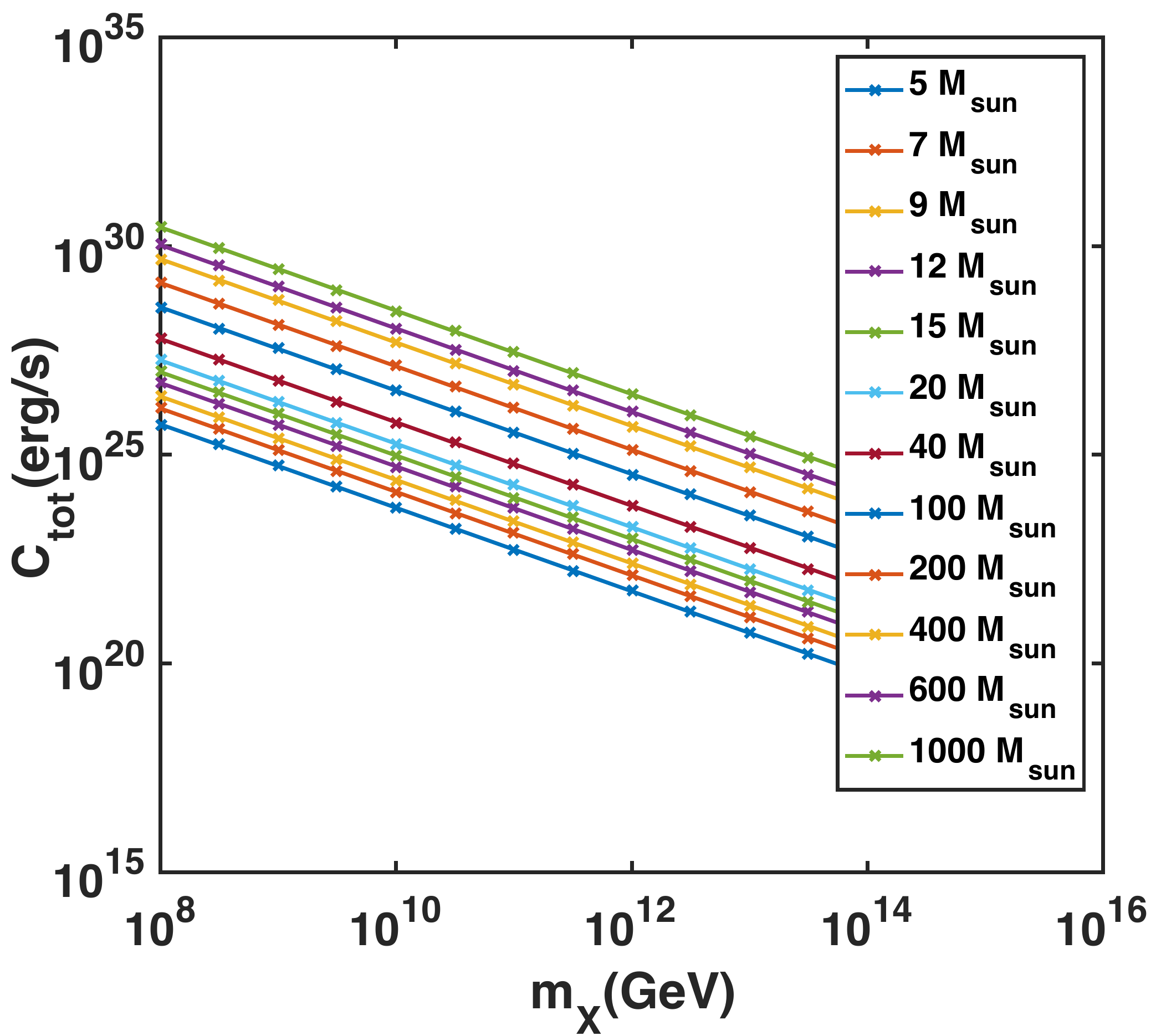}
\caption{\label{fig:CtotvsmX} Upper bounds on the total capture rate for SHDM of mass $10^{8}\unit{GeV} - 10^{15}\unit{GeV}$ by Pop.~III stars of various masses. Note that the higher the mass, the higher the capture rate and that for the entire mass range explored $C_{tot}\propto m_X^{-1}$.}
\end{figure}

We next do two consistency checks, to reproduce results previously published in the literature. First, in the case of WIMP (i.e. $m_X\sim 100\unit{GeV}$), and for constant scattering cross section, the capture can be calculated using the single scattering formalism, and rate scales as: $C\propto 1/m_X$~\cite{Freese:2008cap}. We have numerically checked that the multi scattering formalism reproduces similar values for the capture rates for the same Pop.~III stars as those considered in~\cite{Freese:2008cap} for capture of WIMPs. We additionally checked numerically that for the case of strongly interacting SHDM of~\cite{Albuquerque:2001} our implementation of the multiscattering capture formalism reproduces the results calculated analytically under the $\tau\gg 1$ assumption in~\cite{Albuquerque:2001}. After those consistency checks we proceed to calculate the upper bounds on the total capture rates for SHDM of mass in the $10^{8}\unit{GeV} - 10^{15}\unit{GeV}$ by Pop.~III stars (see figure~\ref{fig:CtotvsmX}). 

It is remarkable that for the entire range of dark matter mass considered, the upper bound on the capture rate scales as $C_{tot}\propto m_X^{-1}$. As one can see from figure~\ref{fig:Ncut}, for all stars considered, at masses below $\sim 10^{12}\unit{GeV}$ the single scattering formalism holds (i.e. $\Ncut=1$). Therefore, as explained in the discussion below equation~\ref{eq:CNscalingsimpl}, $C_1\propto \sigma_n\rho_X/m_X^2$. The upper bound on the scattering cross section scales linearly with $m_X$(see equation~\ref{eq:signbound}). Therefore $C_1\propto \rho_X/m_X$, as evidenced by trend at the lower mass range in figure~\ref{fig:CtotvsmX}. At the higher mass range, when $\tau > 1$ we now have to consider adding multiple terms that each scale, in the large N limit, approximately like: $C_N\propto N^2\rho_X/(\sigma_n m_X)^2$. Note that we have dropped the $\Gamma(N+2,\tau)/N!$ term, since this term is approximately zero whenever $N<\tau$ (i.e. for $N<\Ncut$). Therefore:

\bea\label{eq:Ctotscaling}
C_{tot}&=&\sum^{\Ncut}_1C_N\sim\int^{\Ncut}_1 C(N)dN\\\nonumber
&&\sim\int^{\Ncut}_1 N^2\frac{\rho_X}{\sigma_n^2m_X^2} dN\sim \Ncut^3 \frac{\rho_X}{\sigma_n^2m_X^2}. 
\eea    
For the masses considered ($m_X\gtrsim 10^{12}\unit{GeV}$) we expect $\Ncut\propto \tau\propto\sigma_n$. This conclusion is reinforced from fitting the data in figure~\ref{fig:Ncut}, as alluded to before. Therefore, we find $C_{tot}\propto\sigma_n\rho_X/m_X^2$ in this regime as well. The fact that the two regimes, single scatter ($\tau\lesssim 1$) and multiple scatter ( $\tau\gtrsim 1$), have the same scaling for the total capture rate can be traced back to the fact that $A_N^2\ll 1$ for both cases, and therefore the approximation for the term in the square brackets used when obtaining equation~\ref{eq:CNscalingsimpl} holds for both cases. From equation~\ref{eq:AN} and table~\ref{table:PopIII} we estimate that the transition to $A_N^2\gtrsim 1$ happens somewhere in the $10^4 - 10^6\unit{GeV}$ range for all the Pop.~III stars considered, with the lower bound corresponding to the $5\Msun$ star, and the upper bound to the $1000\Msun$ case.~\footnote{In estimating the $m_X$ values quoted we kept $\overline{v}=10\unit{km/s}$.} Furthermore, for this mass range $\tau \ll 1$, so effectively $N=1$ (i.e. we only have to consider single scattering). When $A_1 \sim 1$ the scaling of $C_1$ and, hence of $C_{tot}$ with the physical parameters of interest is more complicated. At even smaller $m_X$, when $A_N^2 \ll 1$, one recovers the expected scaling for WIMPs: $C_{tot}=C_1\propto \sigma_n\rho_X/m_X$.

We point out that for the case of $A_N^2\ll 1$, relevant to this work, the sum defining the total capture rate can be estimated analytically by substituting $(N+1-\Gamma(N+2,\tau)/N!)$  with $\tau^2 p_N(\tau)/2$ in equation~\ref{eq:CNscalingsimpl} and then noting that
\[
 \sum_{N=1}^{N=\infty}p_N(\tau)N=N_{avg}\simeq\tau 
\]
With this, and using the approximation for $C_N$ from equation~\ref{eq:CNscalingsimpl} we get the following estimates for the total capture rate:
\be\label{eq:CtotApprox}
C_{tot}\simeq10^{24}s^{-1}\left( \dfrac{\sigma_{n}}{1.26\times10^{-40}cm^{2}}\right)\left(\dfrac{\rho_{X}}{10^9 GeV /cm^{3}}\right)\left(\dfrac{10^{8}GeV}{m_{X}}\right)^2 \left( \dfrac{10km/s}{\overline{v}}\right)^3
\left(\frac{\Mstar}{\Msun}\right)^3\left(\dfrac{R_{\odot}}{R_{\star}}\right)^{2}
\ee
For the scaling of the upper bound on the total capture rate, in view of $\sigma_n\propto m_X$ we find, as expected from figure~\ref{fig:CtotvsmX}, $C_{tot}\propto m_X^{-1}$.  It is worth reminding the reader that here when calculating the upper bounds on $C_N$ we assume the scattering cross section at the limit allowed by the XENON1T one year exposure limits for spin independent (SI) DM-nucleon scattering. The upper limits on spin dependent (SD) scattering cross section are weaker by a few orders of magnitude, so if one were to use the SD scattering, our upper bounds on the total capture rate would be enhanced by the same factor.  

We summarize here our most important results so far. For the case of SHDM with masses $m_{X}\gtrsim 10^{12}\unit{GeV}$, when multiscatter capture becomes important, the interplay between the $\tau$ dependence of $\Ncut$ and the scaling of each term in the series with $N$ leads to a  total capture rate  that scales as: $C_{tot}\propto(\sigma_n\rho_X/m_X^2)$. This is one of the most important, and perhaps counter intuitive results, of our work. Additionally, we find that for SHDM with $m_X\lesssim 10^{12}\unit{GeV}$ the multiscatter formalism is redundant, when considering Pop.III stars, and therefore $C_{tot}=C_1\propto\sigma_n\rho_X/m_X^2$. It is remarkable that the total capture rates has the same scaling in both regimes.  Regarding upper bounds of the total capture rate, we note that for the entire SHDM mass range considered ($10^{8} - 10^{15}\unit{GeV}$) we obtain: $C_{tot}\propto\rho_X/m_X$. We will investigate the possible consequences of this finding on the maximum mass of Pop.~III stars in the next section.   

\section{Pop.~III stellar mass limits}
\label{sec:masslimit}
After DM particles get captured by any object, they can self-annihilate, assuming they are their own anti-partners. This could lead to a new source of energy for the star, and potentially could even disrupt its evolution or impose a cutoff on the maximum mass for a Pop.~III star. For the case of WIMPs this was studied by~\cite{Freese:2008cap}. We proceed here to do a similar analysis for the case of SHDM capture. 

The number of dark matter particles in the star can be modeled by the competition between capture and annihilation, described mathematically by the differential equation below:~\footnote{We neglect evaporation effects, as those are only relevant for sub-$\unit{GeV}$ DM particle mass.}
\begin{align}\label{eq:Ndiff}
\dot{N}=C-2\Gamma_{A},
\end{align}
where $C(s^{-1})$ is the capture rate, $ \Gamma_{A}(s^{-1}) $ the annihilation rate. 

\begin{figure}[!htb]
\centering 
\includegraphics[width=0.7\textwidth]{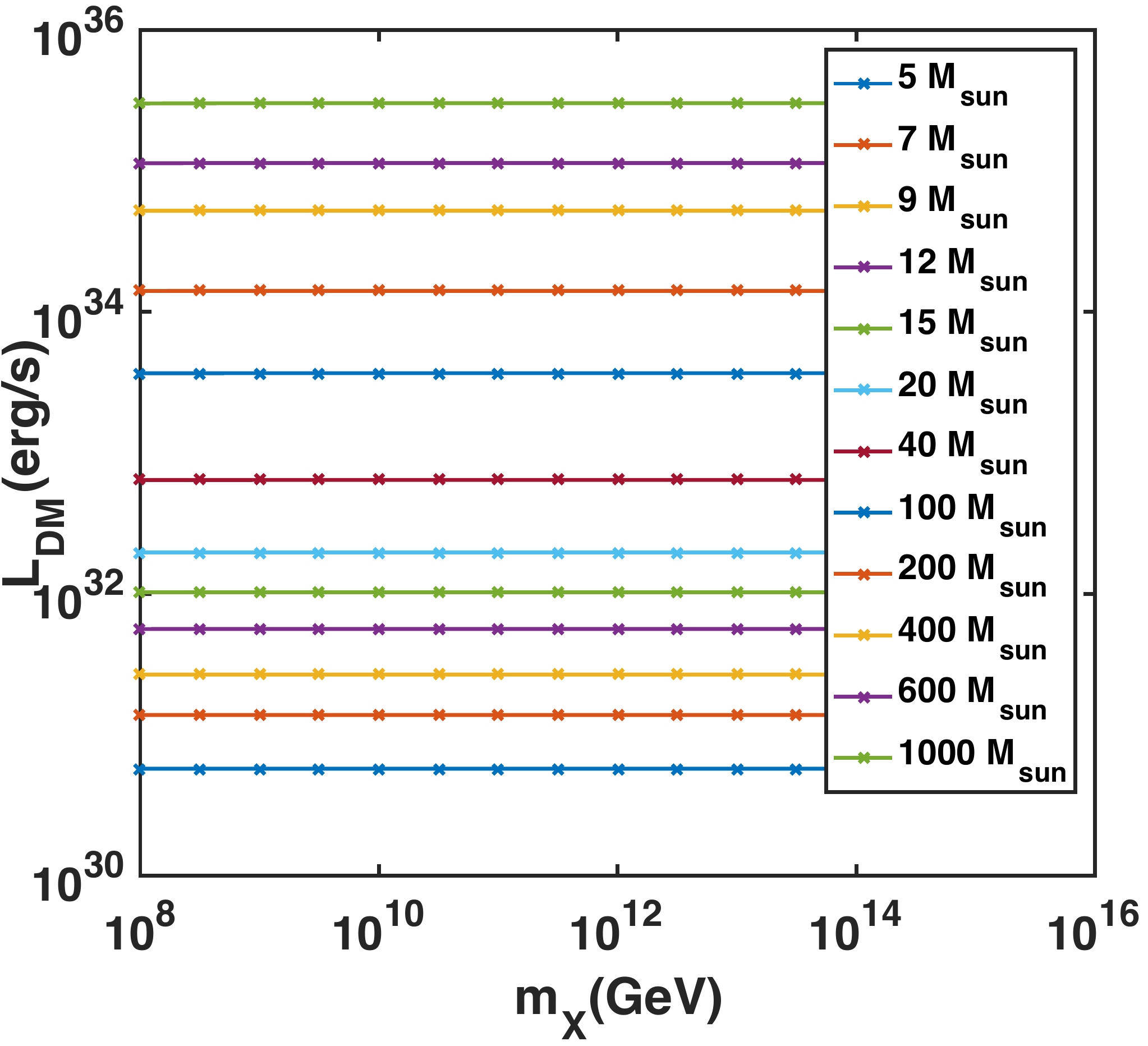}
\caption{\label{fig:LDM}Upper bounds on the luminosity due to DM annihilation for SHDM of mass in the  $10^{8}\unit{GeV} - 10^{15}\unit{GeV}$ range captured by Pop.~III stars of various masses. Note that the upper bounds are insensitive to $m_X$, for the entire range considered.}
\end{figure}

For WIMPs, this equilibrium is quickly reached within the lifetime of the star by a time scale $\tau$~\cite{Freese:2008cap}. We have checked numerically that, assuming a distribution where most of the DM captured is located near the core of the star, this conclusion holds in our case as well. At the equilibrium where the rate of change of the numbers of the dark matter particles in the star equals to zero, one has the following  relationship between capture and annihilation:
\begin{align}\label{18}
\Gamma_{A}=\dfrac{1}{2}C.
\end{align} 
At this time the DM in the core can provide an additional stabilizing energy source, with a total luminosity :

\be\label{eq:LDM}
L_{DM}=f\Gamma_A 2m_X=f C m_X,
\ee
since an energy of $2m_X$ is released after each annihilation event.  Henceforth $f$ represents the efficiency of energy conversion from dark matter particle to power up the star, for which we will assume, following~\cite{Spolyar:2008dark}, a value of $2/3$. This convention amounts to one third of the annihilation energy being lost to neutrinos, and the other two thirds being deposited in the star. In figure~\ref{fig:LDM} we plot the luminosity due to dark matter annihilations, assuming equilibrium between capture an annihilations. We mention in passing that if this assumption is dropped, one would need to develop a Monte Carlo simulation, tracking each DM particle, as it gets captured and annihilated.  

We next proceed to estimate an upper bound on the mass of Pop.~III stars, by using the Eddington limit. As a star becomes purely radiation pressure dominated, its luminosity will become linearly proportional with mass, in what is known as the Eddington limit. For a given stellar mass $M_{\star}$, no star can shine brighter than the Eddington luminosity, as any further accretion of mass is disrupted by the radiation pressure. The Eddington luminosity is defined as:  
\begin{align}\label{20}
L_{Edd}=\dfrac{4\pi c GM_{\star}}{\kappa_{\rho}},
\end{align}
where $G$ is the gravitation constant, $c$ is the speed of light, $M_{\star}$ is the mass of the star and $ \kappa_{\rho} $ is the opacity of the stellar atmosphere. Since the first stars are metal free with hot atmosphere, the opacity is mainly due to Thompson electron scattering. For ease of comparison with previous results from the literature for capture of WIMPs~\cite{Freese:2008cap}, we adopt the same opacity, and hence the same value for the Eddington limit:
\be\label{eq:edd}
L_{Edd}=3.5 \times 10^{4} (M_{\star}/M_{\odot})L_{\odot}.
\ee

In reality, for Thompson electron scattering $\kappa$  depends only on the hydrogen fraction (X) and takes a particularly simple form: $\kappa=\kappa_{es}=0.2(1+X)\unit{cm^2s^{-1}}$. For any stars the value of X will change, as the star burns H into He, and thus even while on the main sequence $X$ is not a constant.  However, for $X$ between $ 1- 0.5$ we get the value of the prefactor in equation~\ref{eq:edd}: ranging between $3.25 - 4.33$, therefore for ZAMS Pop.~III stars, to a good approximation the Eddington limit is represented by the mass dependent value in equation~\ref{eq:edd}.

We use two criteria for finding an approximation for the maximum mass of Pop.~III stars, when including the effects of heating from captured DM-DM annihilations: 
\bea\label{eq:Mmax}
L_{Edd}(\Mmax)&=&L_{nuc}(M_{max})+L_{DM}(M_{max}),\\
L_{Edd}(\Mmax)&=&L_{DM}(M_{max})\label{eq:Mmaxrelaxed} 
\eea
For $L_{nuc}$ we take the values of the stellar luminosities for Pop.~III stars, where no DM heating is included, as tabulated in table~\ref{table:PopIII}. The first criterion will lead to somewhat more stringent bounds; however, it assumes that the rate of hydrogen burning will not be affected by the additional heat source from DM-DM annihilations. The second criterion is very strict, in the sense that it completely disregards any energy from nuclear fusion. In reality, the actual value for $M_{max}$ will be somewhere in between the two bounds, closer to the one obtained using the first criterion. For a more precise estimation one would need to run a stellar evolution code with the effect of DM heating annihilation included. However, for our purposes, where an order of magnitude estimate of the upper limit of the stellar mass is what we are looking for, this approach would be overkill. 

\begin{figure}[!htb]
\centering 
\includegraphics[width=.45\textwidth]{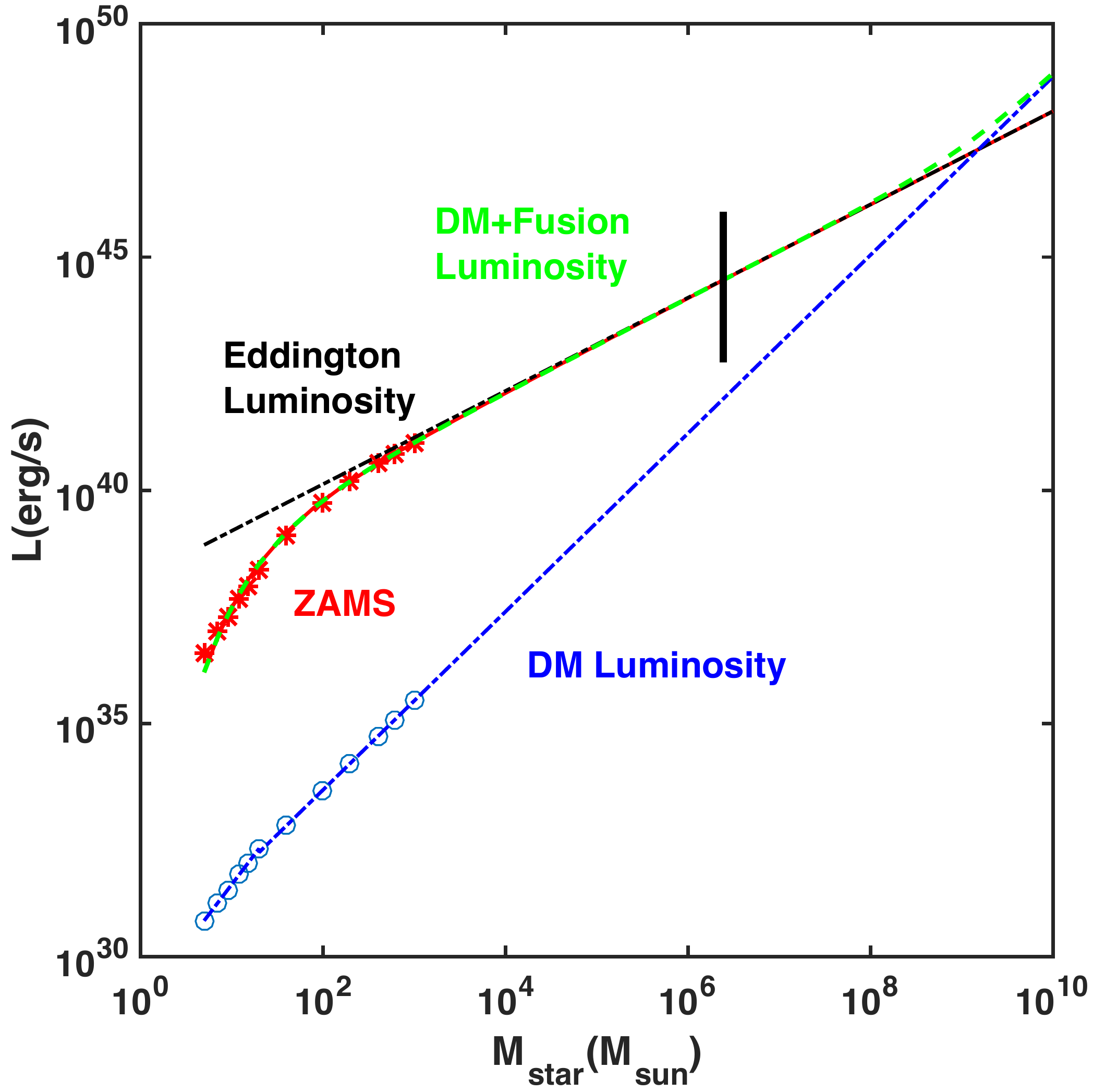}
\hfill
\includegraphics[width=.45\textwidth]{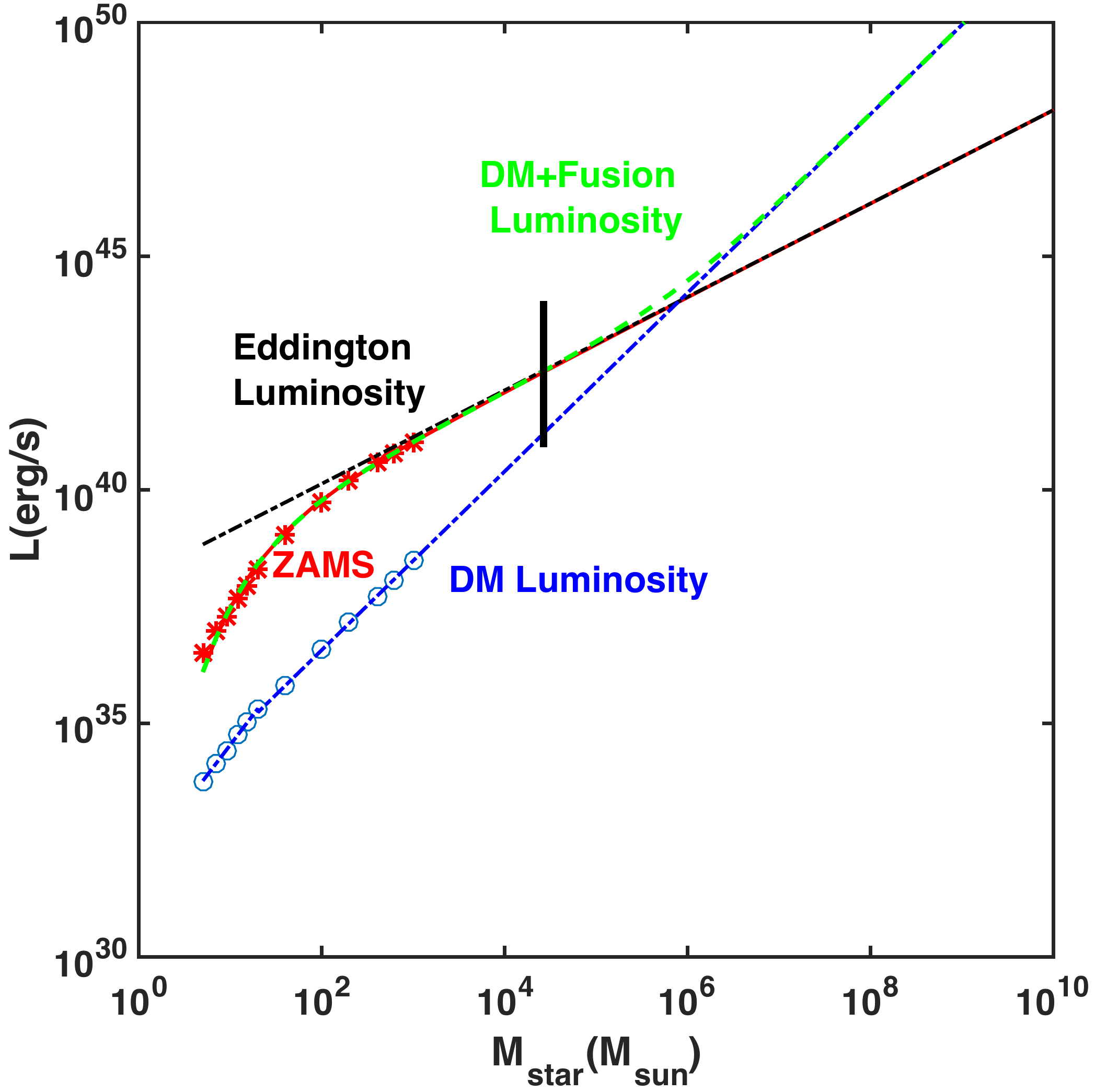}
\caption{\label{fig:Mmax1} Various luminosities as a function of stellar mass. As expected, the luminosity due to nuclear fusion (the solid line interpolating through the stars symbols labeled ZAMS) approaches the Eddington limit $M_{\star}\gtrsim 100\Msun$. The luminosity due to DM annihilations increases faster than linearly, so there will always be a stellar mass beyond which $L_{DM}>L_{Edd}$. The bold vertical line corresponds to the value of $\Mmax$ where $L_{DM}+L_{nuc}>L_{Edd}$. The dark matter ambient density is taken $\rho_X=10^9\unit{GeV/cm^3}$(left panel) and $\rho_X=10^{12}\unit{GeV/cm^3}$ (right panel). Note that an increase of $\rho_X$ leads to a smaller $\Mmax$}
\end{figure}

In figure~\ref{fig:Mmax1} we plot the relevant luminosities as a function of stellar mass. We note that $L_{DM}$ can be very well approximated with a broken power law. In fact, combining the estimate for the total capture rate from equation~\ref{eq:CtotApprox} with the luminosity due to dark matter, when the equilibrium between capture and annihilation has been reached: $L_{DM}=fC_{tot}m_X$, we get:
\bea\nonumber
L_{DM}&\approx& 10^{29}\unit{erg/s}\left( \dfrac{\sigma_{n}}{1.26\times10^{-40}cm^{2}}\right)\left(\dfrac{\rho_{X}}{10^9 GeV /cm^{3}}\right)\left(\dfrac{10^{8}GeV}{m_{X}}\right)\left( \dfrac{10km/s}{\overline{v}}\right)^3\\
&&\left(\frac{\Mstar}{\Msun}\right)^3\left(\dfrac{R_{\odot}}{R_{\star}}\right)^{2}\label{eq:LDMapprox}
\eea
Using the homology scaling relation for the higher mass Pop.~III stars found in equation~\ref{eq:homologyhigh} one gets $L_{DM}\propto \Mstar^{1.88}$. This is a remarkable result and is in contrast to the scaling obtained for WIMP capture by~\cite{Freese:2008cap}, where $L_{WIMP}\propto \Mstar^{1.55}$. For SHDM capture, the increase in the DM luminosity with the mass of the star is significantly faster than the one for WIMPs. As explained before, when discussing the scaling of the total capture rate, the differences between the scaling relationships for WIMPs vs SHDM can be attributed to the $A_N^2\ll 1$ (WIMPs) vs. $A_N^2 \gg 1$ (SHDM), and therefore the different approximations valid for the term in the square brackets of equation~\ref{eq:CNsimpl}.  

For the nuclear fusion luminosity of ZAMS zero metallicity stars (Pop.~III stars) we find the following approximate fitting formula:
\be\label{eq:fitZAMSZ0}
L_{nuc}\simeq 10^\frac{\log(3.5\times10^4L_{\odot}\unit{s/erg})}{1+\exp(-1.03x-1.17)}\cdot x^{\frac{8.27}{x^{0.59}}+1}~\unit{erg/s},
\ee
where $x\equiv\frac{M_{\star}}{\Msun}$ and $L_{\odot}\equiv3.846\times10^{33}\unit{erg/s}$. This interpolates between the expected scaling regimes $L_{nuc}\propto M_{\star}^3$ for lower mass stars, and $L_{nuc}\propto M_{\star}$, for high mass stars, which are radiation pressure dominated and shine at the Eddington limit. 

Since $L_{DM}\propto \rho_X$, an increase in the DM ambient density will lead to a lower value of $\Mmax$. One mechanism identified in the literature that can lead to an increase in $\rho_X$ at the center of a DM halo during the formation of a star is adiabatic contraction~\cite{Blumenthal:1985}. For instance,~\cite{Spolyar:2008dark} shows that during the formation of the first stars adiabatic contraction can lead to an enhancement of $\rho_X$ by many orders of magnitude. If adiabatic contraction operates until the protostellar core gas has a number density $n\sim 10^{22}\unit{cm^{-3}}$, the DM density would reach $\rho_X\sim 10^{18}\unit{GeV/cm^3}$. x
\begin{figure}[!htb]
\centering 
\includegraphics[width=.7\textwidth]{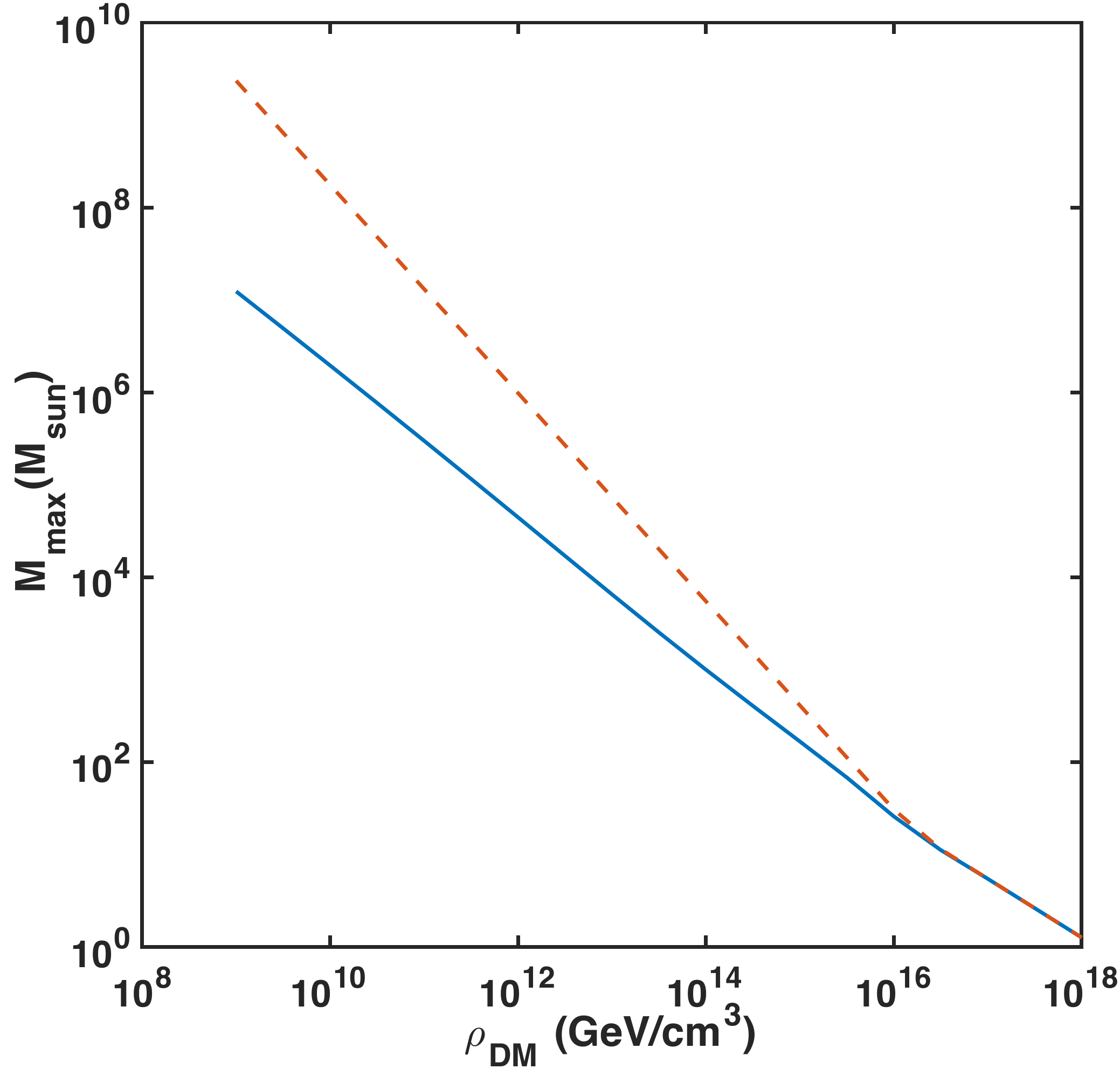}
\caption{\label{fig:Mmaxvsrho} Maximum stellar mass, when including the effects of annihilation of captured SHDM by Pop III stars. For the solid (blue) line the following condition is used: $L_{nuc}(M_{max})+L_{DM}(M_{max})=L_{Edd}(\Mmax)$ to find the value of $\Mmax$. For the dashed (orange) line we used the weaker $L_{DM}(\Mmax)=L_{Edd}(\Mmax)$ condition.}
\end{figure}
In figure~\ref{fig:Mmaxvsrho} we plot the maximum Pop.~III stellar mass versus $\rho_X$. The kink at $\rho_X\approx 10^{16}\unit{GeV/cm^3}$ is due to the different radius-mass homology relations for low vs high mass Pop.~III stars. Note that if the DM ambient density reaches $\rho_X\sim 10^{14}\unit{GeV/cm^3}$ the effects of captured SHDM annihilations are able to prevent stars with masses larger than roughly $1000\Msun$. At lower densities the maximum stellar mass is likely to be controlled by other effects, such as fragmentation of the protostellar gas cloud. However, when $\rho_X\gtrsim 10^{14}\unit{GeV/cm^3}$, SHDM capture can place very stringent bounds on the maximum mass a Pop.~III star can have. For the extreme case of $\rho_X\approx10^{18}\unit{GeV/cm^3}$ the maximum Pop.~III mass becomes roughly one solar mass!  

In the case of the less stringent criterion for $\Mmax$ we find, via equations~\ref{eq:edd} and~\ref{eq:LDMapprox}, the following approximation:
\be
\left(\frac{\Mstar}{\Msun}\right)\left(\frac{\Rsun}{\Rstar}\right)\lesssim3.67\times 10^4\left(\frac{1.26\times10^{-40}cm^2}{\sigma_n}\right)^{1/2}\left(\frac{m_X}{10^8\unit{GeV}}\right)^{1/2}\left(\frac{10^9\unit{GeV/cm^3}}{\rho_X}\right)^{1/2}\left(\frac{\overline{v}}{10\unit{km/s}}\right)^{3/2}\nonumber
\ee
If the DM ambient density is not higher than roughly $10^{16}\unit{GeV/cm^2}$ we can use the mass-radius scaling relations of equation~\ref{eq:homologyhigh}, valid for $\Mstar\gtrsim 20\Msun$. For higher DM densities $\Mmax\lesssim 20\Msun$, so one will need to use the mass-radius scaling relations of equation~\ref{eq:homologylow}. Assuming the cross section given by the upper bound of the XENON1T one year exposure (see equation~\ref{eq:signbound}) we find:
\bea
\Mmax &\approx& 20\Msun\left(\frac{10^{16}\unit{GeV/cm^3}}{\rho_X}\right)^{1/0.88}\left(\frac{\overline{v}}{10\unit{km/s}}\right)^{3/0.88}\,  
\mathrm{ for }\, \rho_X\lesssim 10^{16}\unit{GeV}\\
\Mmax &\approx& 20\Msun\left(\frac{10^{16}\unit{GeV/cm^3}}{\rho_X}\right)^{1/1.58}\left(\frac{\overline{v}}{10\unit{km/s}}\right)^{3/1.58}\,
\mathrm{ for }\, \rho_X\gtrsim 10^{16}\unit{GeV}.
\eea
We confirmed this result using a numerical fit of the relevant points from figure~\ref{fig:Mmaxvsrho}. Including the effect of the nuclear fusion luminosity when calculating $\Mmax$, leads to a milder dependence of $\Mmax$ with $\rho_X$. By fitting with a power law we find $\Mmax\propto\rho_X^{-0.81}$ (when $\rho_X\lesssim 10^{16}\unit{GeV}$) and $\Mmax\propto\rho_X^{-0.63}$ (when $\rho_X\gtrsim 10^{16}\unit{GeV}$). From figure~\ref{fig:Mmaxvsrho} note that the two criteria are essentially indistinguishable at $\rho_X\gtrsim 10^{16}\unit{GeV/cm^3}$. This is due to the fact that at those high DM densities the luminosity due to DM annihilations dominates over the one due to fusion for the stars with masses close to the respective upper bounds.    

In conclusion, annihilation of SHDM captured by Pop.~III stars can lead to a maximum value of the stellar mass that is inversely proportional to the ambient DM density $\rho_X$. This confirms results previously obtained in~\cite{Freese:2008cap} for WIMPs and extends them to the capture of SHDM. If the DM ambient density is as high as $10^{18}GeV/cm^3$, i.e. if we assume adiabatic contraction throughout the entire history of the protostellar gas cloud collapse, we find that heating due to annihilations of captured SHDM prevents Pop.~III stars from having masses larger than roughly a solar mass! We mention in passing that this argument can also be turned around and used to constrain the following combination of dark matter parameters: $(\rho_X\sigma_n/m_X)$, since the dark matter heating is directly proportional to this combination of parameters. Namely, for any future observed Pop.~III star of $M_{obs}$, one can rule out the combination $(\rho_X\sigma_n/m_X)$ that would lead to $L_{DM}>L_{Edd}(M_{obs})$. This opens up the possibility to place constraints on properties of dark matter from the observed masses of the first stars, which should be possible with JWST.   

\nopagebreak
\section{Conclusion}
\label{sec:conclusion}
In this paper we have investigated the capture of SHDM by Pop.~III stars. Using the multi scattering formalism, we find, for SHDM with masses in the $10^{8} - 10^{16}\unit{GeV}$ range, that the total capture rate has the following scaling with the relevant parameters: $(\rho_X\sigma_n/m_X^2)(M_{\star}^3/\Rstar^2)$. This leads to an upper bound on the capture rate that is proportional with $1/m_X$, which, in turn implies a $m_X$ independent upper bound on the luminosity due to captured SHDM annihilations. We also show that if a non-zero fraction of the annihilation products can thermalize and deposit energy inside the star, the masses of Pop. III stars have an upper bound ($\Mmax$) that is inversely proportional to the ambient dark matter density. We estimate values for $\Mmax$ for ambient DM densities ranging from $10^9- 10^{18}\unit{GeV/cm^3}$. At $\rho_X\sim 10^{16}\unit{GeV/cm^3}$, and assuming the scattering cross section given by the XENON1T one year detection bounds for one year exposure, we obtain a maximum value of the mass of a Pop.~III star of roughly a few tens of solar masses. In the extreme case of $\rho_X\sim 10^{18}\unit{GeV/cm^3}$ Pop.~III stars could not grow much past a few solar masses, in view of the Eddington limit. The existence of a dark matter dependent upper limit on Pop.~III stellar masses raises the possibility to constrain properties of dark matter by using upcoming observations of first stars with JWST, and their inferred masses.  

\appendix
\section{Captured SHDM Annihilations}
\label{sec:dmannih}

Dark Matter captured by any star will settle to the core on a time scale that is determined by their drift velocity. Specifically, $\tau_{drift}\sim r/v_{drift}$, with the drift velocity being found by imposing the equilibrium between the gravitational force and the viscous drag force due to scatterings with nucleons~\cite{Gould:1989,Albuquerque:2001}: 
\be\label{eq:viscous}
G\bar{\rho}r^3m_X/r^2\sim \sigma_N n_N v_{drift}(r)m_{N}\left(k_B\bar{T}/m_N\right)^{1/2},
\ee
with $\bar{\rho}$ and $\bar{T}$ being the typical mass density and temperature of the star. Since we are only interested in an order of magnitude estimate we can set $\bar{\rho}\sim n_N m_N$, leading to $\tau\propto \sigma_N m_X^{-1}(\bar{T})^{1/2}$. For all the DM masses considered in our paper $\tau_{drift}$ is very short, much less than the lifetime of the star. Therefore, quickly after being captured SHDM particles sink to the core of the Pop.~III star, where they could annihilate.

We next detail the steps taken to estimate the time scale after which an equilibrium between capture of supermassive dark matter by a compact object and its subsequent annihilations is attained. We are following the same procedure as~\cite{Freese:2008cap} did for the case of WIMPs being captured. The main difference comes from the annihilation cross section, which for thermal relics, such as WIMPs, is set by the dark matter relic abundance. For the case of non-thermal relics, such as the SHDM considered in this paper, the annihilation cross section is not tied to the relic abundance. However, the unitarity bound restricts the annihilation cross section:

\be\label{eq:unitarity}
\sigmav\leq\frac{4\pi\hbar^2}{m_{X}^2v}\lesssim 4.4\times10^{-37}\frac{\unit{cm}^3}{\unit{s}}\left(\frac{10^{12}\unit{GeV}}{m_X}\right)^2\left(\frac{100km/s}{v}\right)
\ee

The number of dark matter particles in the star can be modeled by the competition between capture and annihilation, described mathematically by the differential equation below:
\begin{equation}
\dot{N}=C-2 \Gamma_{A} \equiv C-C_{A} N^{2}
\end{equation}
where $C(s^{-1}) $ is the capture rate, and $\Gamma_{A}(s^{-1})$ is the annihilation rate,
\begin{equation}
C_{A}=2 \Gamma_{A} / N^{2}
\end{equation}
is defined as an $N-$independent annihilation coefficient. Solving this equation, we find
\begin{equation}\label{eq:tau}
\Gamma_{A}=\frac{1}{2} C \tanh ^{2}(t / \tau),\,\mathrm{with}\,\,\tau=\left(C C_{A}\right)^{-1 / 2},
\end{equation}
At the equilibrium between annihilations and capture, when the rate of change of the numbers of the dark matter particles in the star equals to zero, one has the following relationship between capture and annihilation:
\begin{equation}
\Gamma_{A}=\frac{1}{2} C.
\end{equation}
Note that for $t\gtrsim\tau$, the annihilation rate of equation~\ref{eq:tau} becomes time independent, as the hyperbolic tangent approaches unity and $\Gamma_A\to 1/2 C$. Therefore $\tau$ corresponds to the time scale for annihilation/capture equilibrium. 

In general, the annihilation rate is given by the following integral over the volume of the star: $\Gamma_A=\int\,dV n^2_{X}(r)\sigmav$. In view of our discussion following equation~\ref{eq:viscous} we can assume that most of the DM captured is located near the core of the star. Therefore, the DM density profile can be approximated using isothermal distribution
\begin{equation}
n_{X}(r)=n_{c} e^{-m_{X} \phi / k T_c}
\end{equation}
where $n_{c}$ the central number density of DM and $T_{c}$ is the central temperature of the star,
\begin{equation}
\phi(r)=\int_{0}^{r} \frac{G M(r)}{r^2} d r
\end{equation}
is the gravitational potential at radius $r$ with respect to the center, and $M(r)$ is the mass interior to $r$. For order of magnitude estimates we assume a constant density profile and therefore $M(r)\sim \rho_c4\pi/3 r^3$, which leads to the following form for the distribution of DM particles~\cite{Griest:1986}: $n_X(r)\propto\exp(-r^2/2r_X^2)$. Note that $r_X$ roughly determines the size of the core within which the capture DM particle are concentrated and it can be expressed as:
\be\label{eq:DMCore}
r_X\equiv\left(\frac{3T_c}{4\pi G m_X \bar{\rho}}\right)^{1/2}
\ee
Inserting numerical values we can estimate the following scaling relation:
\be
r_X=5\times 10^3\unit{cm}\left(\frac{10^{12}~\unit{GeV}}{m_X}\right)^{1/2}\left(\frac{T_c}{10^7 K}\right)^{1/2}\left(\frac{150~\unit{g}\unit{cm}^{-3}}{\rho_c}\right)^{1/2},
\ee
which shows the size of the DM core within which annihilations could happen efficiently is very small, compared to the full radius of the star. This holds for the entire range of parameters in both DM mass and stellar temperature and densities considered here.  
 
 We next proceed to calculate the time scale $\tau$ that sets the equilibrium between capture and annihilations. One can define effective volumes
\begin{equation}
V_{j}=4 \pi \int_{0}^{R_{*}} r^{2} e^{-j m_{X} \phi / T_{c}} \mathrm{d} r =\left[3 m_{\mathrm{pl}}^{2} T_{c} /\left(2 j m_{X} \rho_{c}\right)\right]^{3 / 2},
\end{equation}
where $m_{pl}$ is the Planck mass, and $\rho_{c}$ is the core mass density of the star. The name “effective volume” is suggestive since we have $N = n_{0}V_{1}$. The total annihilation rate is given as
\begin{equation}
\Gamma_{A}=\int d^{3} r n_{X}(r)^{2}\langle\sigma v\rangle_{\mathrm{ann}}=\langle\sigma v\rangle_{\mathrm{ann}} n_{o}^{2} V_{2},
\end{equation}
where the annihilation cross section
\begin{equation}
\langle\sigma v\rangle_{\mathrm{ann}} = \frac{4 \pi \hbar^{2} c^{4}}{m_{X}^{2} v_{r e l}},\,\, v_{rel}=\sqrt{\dfrac{3kT_{c}}{m_{X}}}
\end{equation} 
is assumed to be at the unitarity bound. The $N-$independent annihilation coefficient becomes:
\begin{equation}
C_{A}=\langle\sigma v\rangle_{\operatorname{ann}} \frac{V_{2}}{V_{1}^{2}}.
\end{equation}
With this  expression for $C_A$ we evaluate the equilibrium timescale $\tau$, defined in equation~\ref{eq:tau}. For the capture rate $C$ we used our approximation from equation~\ref{eq:CtotApprox}. For the stellar and dark matter parameters considered in our work this time scale ranges from $10^{-3}\unit{yr}$ to $10\unit{yrs}$. This is well within the lifetime of the star, and therefore we can assume an equilibrium between capture and annihilation rates. Therefore, the luminosity due to annihilations of captured dark matter can be related to the capture rate: $L_{DM}=f\Gamma_A2m_X=fC m_X$, as we did in section~\ref{sec:masslimit}.

\bibliographystyle{JHEP}
\bibliography{RefsDM}

\providecommand{\href}[2]{#2}\begingroup\raggedright\begin{thebibliography}{10}

\bibitem{Barkana:2000}
R.~Barkana and A.~Loeb, \emph{{In the beginning: The First sources of light and
  the reionization of the Universe}},
  \href{https://doi.org/10.1016/S0370-1573(01)00019-9}{\emph{Phys. Rept.}
  {\bfseries 349} (2001) 125}
  [\href{https://arxiv.org/abs/astro-ph/0010468}{{\ttfamily
  astro-ph/0010468}}].

\bibitem{Abel:2001}
T.~Abel, G.~L. Bryan and M.~L. Norman, \emph{{The formation of the first star
  in the Universe}}, \href{https://doi.org/10.1126/science.295.5552.93,
  10.1126/science.1063991}{\emph{Science} {\bfseries 295} (2002) 93}
  [\href{https://arxiv.org/abs/astro-ph/0112088}{{\ttfamily
  astro-ph/0112088}}].

\bibitem{Bromm:2003}
V.~Bromm and R.~B. Larson, \emph{{The First stars}},
  \href{https://doi.org/10.1146/annurev.astro.42.053102.134034}{\emph{Ann. Rev.
  Astron. Astrophys.} {\bfseries 42} (2004) 79}
  [\href{https://arxiv.org/abs/astro-ph/0311019}{{\ttfamily
  astro-ph/0311019}}].

\bibitem{Yoshida:2006}
N.~Yoshida, K.~Omukai, L.~Hernquist and T.~Abel, \emph{{Formation of Primordial
  Stars in a lambda-CDM Universe}},
  \href{https://doi.org/10.1086/507978}{\emph{Astrophys. J.} {\bfseries 652}
  (2006) 6} [\href{https://arxiv.org/abs/astro-ph/0606106}{{\ttfamily
  astro-ph/0606106}}].

\bibitem{Yoshida:2008}
N.~Yoshida, K.~Omukai and L.~Hernquist, \emph{{Protostar Formation in the Early
  Universe}}, \href{https://doi.org/10.1126/science.1160259}{\emph{Science}
  {\bfseries 321} (2008) 669}
  [\href{https://arxiv.org/abs/0807.4928}{{\ttfamily 0807.4928}}].

\bibitem{Loeb:2010}
A.~Loeb, \emph{{How did the first stars and galaxies form?}} Princeton
  University Press, Princeton, NJ, 2010.

\bibitem{Bromm:2013}
V.~{Bromm}, \emph{{Formation of the first stars}},
  \href{https://doi.org/10.1088/0034-4885/76/11/112901}{\emph{Reports on
  Progress in Physics} {\bfseries 76} (2013) 112901}
  [\href{https://arxiv.org/abs/1305.5178}{{\ttfamily 1305.5178}}].

\bibitem{Klessen:2018}
R.~S. {Klessen}, \emph{{Formation of the first stars}}, {\emph{arXiv e-prints}
  (2018) arXiv:1807.06248} [\href{https://arxiv.org/abs/1807.06248}{{\ttfamily
  1807.06248}}].

\bibitem{Spolyar:2008dark}
D.~Spolyar, K.~Freese and P.~Gondolo, \emph{{Dark matter and the first stars: a
  new phase of stellar evolution}},
  \href{https://doi.org/10.1103/PhysRevLett.100.051101}{\emph{Phys. Rev. Lett.}
  {\bfseries 100} (2008) 051101}
  [\href{https://arxiv.org/abs/0705.0521}{{\ttfamily 0705.0521}}].

\bibitem{Iocco:2008}
F.~{Iocco}, A.~{Bressan}, E.~{Ripamonti}, R.~{Schneider}, A.~{Ferrara} and
  P.~{Marigo}, \emph{{Dark matter annihilation effects on the first stars}},
  \href{https://doi.org/10.1111/j.1365-2966.2008.13853.x}{\emph{\mnras}
  {\bfseries 390} (2008) 1655}
  [\href{https://arxiv.org/abs/0805.4016}{{\ttfamily 0805.4016}}].

\bibitem{Blumenthal:1985}
G.~R. Blumenthal, S.~M. Faber, R.~Flores and J.~R. Primack, \emph{{Contraction
  of Dark Matter Galactic Halos Due to Baryonic Infall}},
  \href{https://doi.org/10.1086/163867}{\emph{Astrophys. J.} {\bfseries 301}
  (1986) 27}.

\bibitem{Press:1985}
W.~H. Press and D.~N. Spergel, \emph{{Capture by the sun of a galactic
  population of weakly interacting massive particles}},
  \href{https://doi.org/10.1086/163485}{\emph{Astrophys. J.} {\bfseries 296}
  (1985) 679}.

\bibitem{Gould:1987}
A.~Gould, \emph{{Direct and Indirect Capture of Wimps by the Earth}},
  \href{https://doi.org/10.1086/166347}{\emph{Astrophys. J.} {\bfseries 328}
  (1988) 919}.

\bibitem{Gould:1987resonant}
A.~{Gould}, \emph{{Resonant enhancements in weakly interacting massive particle
  capture by the earth}}, \href{https://doi.org/10.1086/165653}{\emph{\apj}
  {\bfseries 321} (1987) 571}.

\bibitem{Freese:2008cap}
K.~Freese, D.~Spolyar and A.~Aguirre, \emph{{Dark Matter Capture in the first
  star: a Power source and a limit on Stellar Mass}},
  \href{https://doi.org/10.1088/1475-7516/2008/11/014}{\emph{JCAP} {\bfseries
  0811} (2008) 014} [\href{https://arxiv.org/abs/0802.1724}{{\ttfamily
  0802.1724}}].

\bibitem{Iocco:2008cap}
F.~Iocco, \emph{{Dark Matter Capture and Annihilation on the First Stars:
  Preliminary Estimates}},
  \href{https://doi.org/10.1086/587959}{\emph{Astrophys. J.} {\bfseries 677}
  (2008) L1} [\href{https://arxiv.org/abs/0802.0941}{{\ttfamily 0802.0941}}].

\bibitem{Sivertsson:2011}
S.~{Sivertsson} and P.~{Gondolo}, \emph{{The WIMP Capture Process for Dark
  Stars in the Early Universe}},
  \href{https://doi.org/10.1088/0004-637X/729/1/51}{\emph{\apj} {\bfseries 729}
  (2011) 51} [\href{https://arxiv.org/abs/1006.0025}{{\ttfamily 1006.0025}}].

\bibitem{Bertone:2008}
G.~Bertone and M.~Fairbairn, \emph{Compact stars as dark matter probes},
  \href{https://doi.org/10.1103/PhysRevD.77.043515}{\emph{Phys. Rev. D}
  {\bfseries 77} (2008) 043515}.

\bibitem{Albuquerque:2001}
I.~F.~M. {Albuquerque}, L.~{Hui} and E.~W. {Kolb}, \emph{{High energy neutrinos
  from superheavy dark matter annihilation}},
  \href{https://doi.org/10.1103/PhysRevD.64.083504}{\emph{\prd} {\bfseries 64}
  (2001) 083504} [\href{https://arxiv.org/abs/hep-ph/0009017}{{\ttfamily
  hep-ph/0009017}}].

\bibitem{Freese:2008ds}
K.~Freese, P.~Bodenheimer, D.~Spolyar and P.~Gondolo, \emph{{Stellar Structure
  of Dark Stars: a first phase of Stellar Evolution due to Dark Matter
  Annihilation}}, \href{https://doi.org/10.1086/592685}{\emph{Astrophys. J.}
  {\bfseries 685} (2008) L101}
  [\href{https://arxiv.org/abs/0806.0617}{{\ttfamily 0806.0617}}].

\bibitem{Spolyar:2009}
D.~Spolyar, P.~Bodenheimer, K.~Freese and P.~Gondolo, \emph{{Dark Stars: a new
  look at the First Stars in the Universe}},
  \href{https://doi.org/10.1088/0004-637X/705/1/1031}{\emph{Astrophys. J.}
  {\bfseries 705} (2009) 1031}
  [\href{https://arxiv.org/abs/0903.3070}{{\ttfamily 0903.3070}}].

\bibitem{Bardeen:1986}
J.~M. {Bardeen}, J.~R. {Bond}, N.~{Kaiser} and A.~S. {Szalay}, \emph{{The
  statistics of peaks of Gaussian random fields}},
  \href{https://doi.org/10.1086/164143}{\emph{\apj} {\bfseries 304} (1986) 15}.

\bibitem{Barnes:1987}
J.~{Barnes} and G.~{Efstathiou}, \emph{{Angular momentum from tidal torques}},
  \href{https://doi.org/10.1086/165480}{\emph{\apj} {\bfseries 319} (1987)
  575}.

\bibitem{Frenk:1988}
C.~S. {Frenk}, S.~D.~M. {White}, M.~{Davis} and G.~{Efstathiou}, \emph{{The
  formation of dark halos in a universe dominated by cold dark matter}},
  \href{https://doi.org/10.1086/166213}{\emph{\apj} {\bfseries 327} (1988)
  507}.

\bibitem{Dubinksi:1991}
J.~{Dubinski} and R.~G. {Carlberg}, \emph{{The structure of cold dark matter
  halos}}, \href{https://doi.org/10.1086/170451}{\emph{\apj} {\bfseries 378}
  (1991) 496}.

\bibitem{Gerhard:1985}
O.~E. {Gerhard} and J.~{Binney}, \emph{{Triaxial galaxies containing massive
  black holes or central density cusps}},
  \href{https://doi.org/10.1093/mnras/216.2.467}{\emph{\mnras} {\bfseries 216}
  (1985) 467}.

\bibitem{Meritt:1996}
D.~{Merritt} and T.~{Fridman}, \emph{{Triaxial Galaxies with Cusps}},
  \href{https://doi.org/10.1086/176957}{\emph{\apj} {\bfseries 460} (1996) 136}
  [\href{https://arxiv.org/abs/astro-ph/9511021}{{\ttfamily
  astro-ph/9511021}}].

\bibitem{Meritt:1996b}
D.~{Merritt} and M.~{Valluri}, \emph{{Chaos and Mixing in Triaxial Stellar
  Systems}}, \href{https://doi.org/10.1086/177955}{\emph{\apj} {\bfseries 471}
  (1996) 82} [\href{https://arxiv.org/abs/astro-ph/9602079}{{\ttfamily
  astro-ph/9602079}}].

\bibitem{Freese:2010smds}
K.~Freese, C.~Ilie, D.~Spolyar, M.~Valluri and P.~Bodenheimer,
  \emph{{Supermassive Dark Stars: Detectable in JWST}},
  \href{https://doi.org/10.1088/0004-637X/716/2/1397}{\emph{Astrophys. J.}
  {\bfseries 716} (2010) 1397}
  [\href{https://arxiv.org/abs/1002.2233}{{\ttfamily 1002.2233}}].

\bibitem{Ilie:2012}
C.~{Ilie}, K.~{Freese}, M.~{Valluri}, I.~T. {Iliev} and P.~R. {Shapiro},
  \emph{{Observing supermassive dark stars with James Webb Space Telescope}},
  \href{https://doi.org/10.1111/j.1365-2966.2012.20760.x}{\emph{\mnras}
  {\bfseries 422} (2012) 2164}
  [\href{https://arxiv.org/abs/1110.6202}{{\ttfamily 1110.6202}}].

\bibitem{Banados:2018}
E.~{Ba{\~n}ados}, B.~P. {Venemans}, C.~{Mazzucchelli}, E.~P. {Farina},
  F.~{Walter}, F.~{Wang} et~al., \emph{{An 800-million-solar-mass black hole in
  a significantly neutral Universe at a redshift of 7.5}},
  \href{https://doi.org/10.1038/nature25180}{\emph{\nat} {\bfseries 553} (2018)
  473} [\href{https://arxiv.org/abs/1712.01860}{{\ttfamily 1712.01860}}].

\bibitem{Banik:2019}
N.~{Banik}, J.~C. {Tan} and P.~{Monaco}, \emph{{The formation of supermassive
  black holes from Population III.1 seeds. I. Cosmic formation histories and
  clustering properties}},
  \href{https://doi.org/10.1093/mnras/sty3298}{\emph{\mnras} {\bfseries 483}
  (2019) 3592} [\href{https://arxiv.org/abs/1608.04421}{{\ttfamily
  1608.04421}}].

\bibitem{Gondolo:2010dmds}
P.~{Gondolo}, J.-H. {Huh}, H.~{Do Kim} and S.~{Scopel}, \emph{{Dark matter that
  can form dark stars}},
  \href{https://doi.org/10.1088/1475-7516/2010/07/026}{\emph{\jcap} {\bfseries
  2010} (2010) 026} [\href{https://arxiv.org/abs/1004.1258}{{\ttfamily
  1004.1258}}].

\bibitem{Duffy:2009}
L.~D. Duffy and K.~van Bibber, \emph{Axions as dark matter particles},
  \href{https://doi.org/10.1088/1367-2630/11/10/105008}{\emph{New Journal of
  Physics} {\bfseries 11} (2009) 105008}.

\bibitem{Kolb:1999}
E.~W. {Kolb}, D.~J.~H. {Chung} and A.~{Riotto}, \emph{{WIMPZILLAS!}},  in
  \emph{Dark matter in Astrophysics and Particle Physics}, H.~V.
  {Klapdor-Kleingrothaus} and L.~{Baudis}, eds., p.~592, Jan, 1999,
  \href{https://arxiv.org/abs/hep-ph/9810361}{{\ttfamily hep-ph/9810361}}.

\bibitem{Bertone:2005}
G.~Bertone, D.~Hooper and J.~Silk, \emph{Particle dark matter: evidence,
  candidates and constraints},
  \href{https://doi.org/https://doi.org/10.1016/j.physrep.2004.08.031}{\emph{Physics
  Reports} {\bfseries 405} (2005) 279 }.

\bibitem{Abdallah:2015}
J.~{Abdallah}, H.~{Araujo}, A.~{Arbey}, A.~{Ashkenazi}, A.~{Belyaev},
  J.~{Berger} et~al., \emph{{Simplified models for dark matter searches at the
  LHC}}, \href{https://doi.org/10.1016/j.dark.2015.08.001}{\emph{Physics of the
  Dark Universe} {\bfseries 9} (2015) 8}
  [\href{https://arxiv.org/abs/1506.03116}{{\ttfamily 1506.03116}}].

\bibitem{Freese:2017dm}
K.~Freese, \emph{{Status of Dark Matter in the Universe}},
  \href{https://doi.org/10.1142/S0218271817300129,
  10.1142/9789813226609_0018}{\emph{Int. J. Mod. Phys.} {\bfseries 1} (2017)
  325} [\href{https://arxiv.org/abs/1701.01840}{{\ttfamily 1701.01840}}].

\bibitem{Bramante:2017}
J.~{Bramante}, A.~{Delgado} and A.~{Martin}, \emph{{Multiscatter stellar
  capture of dark matter}},
  \href{https://doi.org/10.1103/PhysRevD.96.063002}{\emph{\prd} {\bfseries 96}
  (2017) 063002} [\href{https://arxiv.org/abs/1703.04043}{{\ttfamily
  1703.04043}}].

\bibitem{Kavanagh:2018}
B.~J. {Kavanagh}, \emph{{Earth scattering of superheavy dark matter: Updated
  constraints from detectors old and new}},
  \href{https://doi.org/10.1103/PhysRevD.97.123013}{\emph{\prd} {\bfseries 97}
  (2018) 123013} [\href{https://arxiv.org/abs/1712.04901}{{\ttfamily
  1712.04901}}].

\bibitem{Aprilie:2018}
{\scshape XENON Collaboration 7} collaboration, \emph{Dark matter search
  results from a one ton-year exposure of xenon1t},
  \href{https://doi.org/10.1103/PhysRevLett.121.111302}{\emph{Phys. Rev. Lett.}
  {\bfseries 121} (2018) 111302}.

\bibitem{Aprilie:2019}
{\scshape XENON Collaboration 4} collaboration, \emph{Constraining the
  spin-dependent wimp-nucleon cross sections with xenon1t},
  \href{https://doi.org/10.1103/PhysRevLett.122.141301}{\emph{Phys. Rev. Lett.}
  {\bfseries 122} (2019) 141301}.

\bibitem{Albuquerque:2010closing}
I.~F.~M. Albuquerque and C.~Perez de~los Heros, \emph{{Closing the Window on
  Strongly Interacting Dark Matter with IceCube}},
  \href{https://doi.org/10.1103/PhysRevD.81.063510}{\emph{Phys. Rev.}
  {\bfseries D81} (2010) 063510}
  [\href{https://arxiv.org/abs/1001.1381}{{\ttfamily 1001.1381}}].

\bibitem{Navarro:1997}
J.~F. {Navarro}, C.~S. {Frenk} and S.~D.~M. {White}, \emph{{A Universal Density
  Profile from Hierarchical Clustering}},
  \href{https://doi.org/10.1086/304888}{\emph{\apj} {\bfseries 490} (1997) 493}
  [\href{https://arxiv.org/abs/astro-ph/9611107}{{\ttfamily
  astro-ph/9611107}}].

\bibitem{Bromm:2001}
V.~{Bromm}, R.~P. {Kudritzki} and A.~{Loeb}, \emph{{Generic Spectrum and
  Ionization Efficiency of a Heavy Initial Mass Function for the First Stars}},
  \href{https://doi.org/10.1086/320549}{\emph{\apj} {\bfseries 552} (2001) 464}
  [\href{https://arxiv.org/abs/astro-ph/0007248}{{\ttfamily
  astro-ph/0007248}}].

\bibitem{Jungman:1996}
G.~{Jungman}, M.~{Kamionkowski} and K.~{Griest}, \emph{{Supersymmetric dark
  matter}}, \href{https://doi.org/10.1016/0370-1573(95)00058-5}{\emph{\physrep}
  {\bfseries 267} (1996) 195}
  [\href{https://arxiv.org/abs/hep-ph/9506380}{{\ttfamily hep-ph/9506380}}].

\bibitem{Ellis:2000}
J.~R. Ellis, A.~Ferstl and K.~A. Olive, \emph{{Reevaluation of the elastic
  scattering of supersymmetric dark matter}},
  \href{https://doi.org/10.1016/S0370-2693(00)00459-7}{\emph{Phys. Lett.}
  {\bfseries B481} (2000) 304}
  [\href{https://arxiv.org/abs/hep-ph/0001005}{{\ttfamily hep-ph/0001005}}].

\bibitem{Cheng:2002}
H.-C. {Cheng}, J.~L. {Feng} and K.~T. {Matchev}, \emph{{Kaluza-Klein Dark
  Matter}}, \href{https://doi.org/10.1103/PhysRevLett.89.211301}{\emph{\prl}
  {\bfseries 89} (2002) 211301}
  [\href{https://arxiv.org/abs/hep-ph/0207125}{{\ttfamily hep-ph/0207125}}].

\bibitem{Schaerer:2002}
D.~{Schaerer}, \emph{{On the properties of massive Population III stars and
  metal-free stellar populations}},
  \href{https://doi.org/10.1051/0004-6361:20011619}{\emph{\aap} {\bfseries 382}
  (2002) 28} [\href{https://arxiv.org/abs/astro-ph/0110697}{{\ttfamily
  astro-ph/0110697}}].

\bibitem{Ohkubo:2009}
T.~{Ohkubo}, K.~{Nomoto}, H.~{Umeda}, N.~{Yoshida} and S.~{Tsuruta},
  \emph{{Evolution of Very Massive Population III Stars with Mass Accretion
  from Pre-main Sequence to Collapse}},
  \href{https://doi.org/10.1088/0004-637X/706/2/1184}{\emph{\apj} {\bfseries
  706} (2009) 1184} [\href{https://arxiv.org/abs/0902.4573}{{\ttfamily
  0902.4573}}].

\bibitem{Windhorst:2019}
R.~{Windhorst}, M.~{Alpaslan}, S.~{Andrews}, T.~{Ashcraft}, T.~{Broadhurst},
  D.~{Coe} et~al., \emph{{On the observability of individual Population III
  stars and their stellar-mass black hole accretion disks through cluster
  caustic transits}}, {\emph{\baas} {\bfseries 51} (2019) 449}
  [\href{https://arxiv.org/abs/1903.06527}{{\ttfamily 1903.06527}}].

\bibitem{Gould:1989}
A.~Gould, B.~T. Draine, R.~W. Romani and S.~Nussinov, \emph{{Neutron Stars:
  Graveyard of Charged Dark Matter}},
  \href{https://doi.org/10.1016/0370-2693(90)91745-W}{\emph{Phys. Lett.}
  {\bfseries B238} (1990) 337}.

\bibitem{Griest:1986}
K.~Griest and D.~Seckel, \emph{{Cosmic Asymmetry, Neutrinos and the Sun}},
  \href{https://doi.org/10.1016/0550-3213(87)90293-8,
  10.1016/0550-3213(88)90409-9}{\emph{Nucl. Phys.} {\bfseries B283} (1987)
  681}.

\end{thebibliography}\endgroup

\end{document}